\documentclass[a4paper, twocolumn, 10pt, superscriptaddress ]{revtex4-1}

\usepackage[utf8]{inputenc}
\usepackage[T1]{fontenc}
\usepackage{lmodern}
\usepackage{ulem}

\usepackage{ifpdf}
\ifpdf
   \usepackage[pdftex]{graphicx}
   \graphicspath{{assets/}}
  \usepackage{epstopdf}
\else
   \usepackage[dvips]{graphicx}
   \graphicspath{{assets/}}
\fi

\usepackage[caption=false,font=footnotesize]{subfig}

\usepackage{bm}

\usepackage{amsmath}
\usepackage{amsthm}
\usepackage{amssymb}
\usepackage{hyperref}
\usepackage{comment}

\usepackage{color}

\newcommand{\jcdbis}[1]{{ #1}}

\newtheorem*{LocalTheo}{Local Time Scale Theorem}

\begin{document}

\title{Severability of mesoscale components and local time scales in dynamical networks}
\author{Yun~William~Yu}
\affiliation{Department of Computer and Mathematical Sciences, University of Toronto, Toronto, ON, Canada}

\author{Jean-Charles~Delvenne}
\affiliation{Institute of Information and Communication Technologies, Electronics and Applied Mathematics, Universit\'{e} catholique de Louvain, Belgium}

\author{Sophia~N.~Yaliraki}
\affiliation{Department of Chemistry, Imperial College London, United Kingdom}

\author{Mauricio~Barahona}
\affiliation{Department of Mathematics, Imperial College London, United Kingdom}

\date{\today}

\begin{abstract}
A major goal of dynamical systems theory is the search for simplified descriptions of the dynamics of a large number of interacting states.
For overwhelmingly complex dynamical systems, the derivation of a reduced description on the entire dynamics at once is computationally infeasible. Other complex systems are so expansive that despite the continual onslaught of new data  only partial information is available.
To address this challenge, we define and optimise for a local quality function \textit{severability} for measuring the dynamical coherency of a set of states over time.
The theoretical underpinnings of severability lie in our local adaptation of the Simon-Ando-Fisher time-scale separation theorem, which formalises the intuition of local wells in the Markov landscape of a dynamical process, or the separation between a microscopic and a macroscopic dynamics.
Finally, we demonstrate the practical relevance of severability by applying it to examples drawn from power networks, image segmentation, social networks, metabolic networks, and word association.
\end{abstract}

\maketitle

Complex dynamical systems composed of a large number of interconnected components are omnipresent, whether in biology (genetic and biochemical networks, interconnected neurons in the brain), technology (power, communication and computer networks) or human interactions (economy and social networks) \cite{majdandzic2014spontaneous,helbing2013globally,watts1998collective, dorfler2011topological, heidemann2012online, ideker2012differential}.
The complexity of such dynamical networks is the result of the subtle interdependence between the dynamics of the individual agents and the network-mediated interactions between them~\cite{strogatz_exploring_2001,dorfler2013synchronization,cornelius2013realistic,watson2005modular}.
As data science becomes pervasive, ever increasing amounts of relational data (in many cases dynamic) are collected and analyzed, and scientists are faced with the challenge of extracting consistent patterns and useful simplified representations in a scalable way.

The search for a simplified description of a complex system that retains essential features of the dynamics of the original is fundamental in complex systems analysis, and many approaches exist, including classical methods such as model reduction for linear systems \cite{moore1981principal,roberts1980linear,pernebo1982model,bultheel1986pade} 
or time scale separation techniques \cite{SimonAndo61,kokotovic86,Coderch83}.
Typically, model reduction (or dimensionality reduction) techniques output a simplified spectral description based on dominant eigenmodes,
which are generally difficult to interpret because their internal states are global combinations of all original states, destroying the original interconnection structure.
Time scale separation techniques are applicable when dynamical systems are dominated by different kinds of behaviour at long and short times.
This coexistence of time scales allows the use of several simplified descriptions, each best suited at a given time scale.
Unfortunately, despite the explosion of data-collection capabilities, many systems are sufficiently complex that only partial data is available.
Often, only the structure of interconnection is available, necessitating a network-centric approach; alternately, sometimes only part of the network is seen or computationally tractable to work with, motivating a local approach.
In this paper, we extend and strengthen the time scale separation approach for complex, heterogeneous network dynamics, focusing on local interactions.

The theory of time scale separation was first explored in detail in the framework of system dynamics by Simon, Ando and Fisher \cite{SimonAndo61,ando1963near}, who considered the existence of a partition of states into components with sufficiently low dynamical influence between them.
At short times, the cross-influence between components can be neglected, so the dynamics of the global system can be approximated by the dynamics of disconnected components.
At long times, the states inside each component evolve to their dominant mode, and the dynamics can be accurately described by the aggregated (or lumped) system, where all the states within each component are collapsed into a single value.
A particularly vivid illustration can be found in Markov chains (or random walks) with coexisting time scales where there are groups of states which mix fast to a quasi-stationary state, yet exhibit low escape probability from each group.
However, the Simon-Ando approach is global in nature, depending on all parameters of the global system---the escape time between groups must be larger than all mixing times to quasi-stationary convergence within each group. Furthermore, the groups of states must be an exact partition of the system, including all nodes and assigning each node to a single group.
Such a uniformity has little reason to emerge spontaneously in large, complex, heterogeneous networks, and can only be artificially imposed by the grouping, splitting or trimming of naturally coherent dynamic structures.

Our work extends the time scale separation approach for complex dynamical systems, and supersedes previous arguments in three ways: it provides a method to detect dynamically coherent structures at all time scales; it
does not require global knowledge of the system; and it allows each state to belong to several overlapping coherent structures or to remain unassigned to any component.
For ease of terminology, we will often use the term `component' to refer to dynamically coherent structures. 
We focus below on Markov random dynamics, or equivalently random walks on weighted networks,
not only because this is an important problem in its own right, but because a significant number of linear and nonlinear dynamics (see Methods) can be reduced to random walks.
Furthermore, random walks admit an intuitive description in terms of flows of probabilities.
In this framework, the \textit{severability} quality function for evaluating a component measures how similarly the full component exchanges probability flows with its surroundings, compared to the component aggregated into a single node. This is measured as a local property of the component, independently from the rest of the system.
In constrast, constructs such as lumped states \cite{derisavi2003optimal}, or macro-states \cite{korenblum2003macrostate} consist in a global partitioning of the states, based on the accuracy of the global reduced dynamics.
The link with network-centric concepts such as clusters \cite{kannan_clusterings:_2004}, communities \cite{girvan_community_2002} is explicited below.

\section{Results}
\subsection{Severability: mixing and retention in Markov landscapes}

\begin{figure*}[tbp]
  \centering
  \includegraphics[width=1\textwidth]{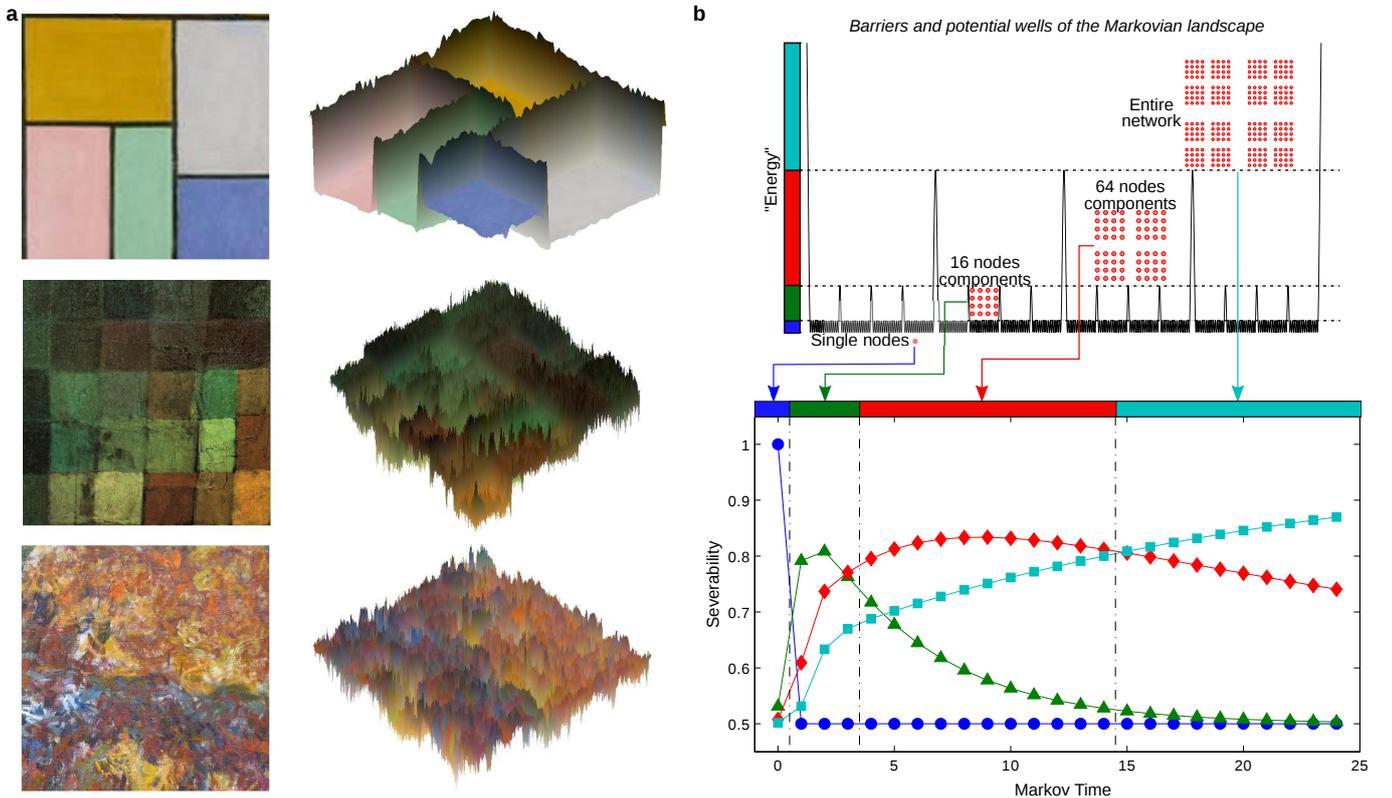}
  \caption{(a)  Left column: small excerpts from three paintings 
      by Theo van Doesburg's \textit{Composition in dissonances} (1919) (top), 
      Paul Klee's \textit{Ancient Sound} (1925) (middle), 
      and Claude Monet's \textit{The Japanese Footbridge} (1920-22) (bottom).
Right column: associated luminosity landscapes obtained from the transition matrix derived
from the graph representation of each image. %
Visualizing the luminosity landscape in van Doesburg's painting reveals coherent spatio-temporal
structures insulated by high barriers, and within which no obstacle would slow down a random walker.
On the other extreme, Monet's rough landscape, when looking at luminosity (i.e. the perceived brightness), is almost featureless with no obvious components.
Klee's landscape is intermediate with significant internal roughness yet noticeable barriers.  %
The balance between the barrier height and the intrinsic roughness translates over to the emergence
of components: barrier heights are inversely related to inter-component connection strength and determine escape time, whereas 
the roughness is inversely related to intra-component connection strength and determines mixing time.
(b)  The ``Markovian'' 
landscape of a hierarchical random graph with three levels and groups of sizes 16, 64, and 256.  
If a pair of nodes are in the same lowest level size 16 component, they are connected with probability $p_1=0.8452$; else if in the same size 64 group, they are connected with probability $p_2=0.0549$; and everything else is connected with probability $p_3=0.036$.
This resulted in an average degree $\langle k \rangle=16$.
The severability of a single node (blue circles), 16-node component (green triangles), 64-node component (red diamond), and the entire network (cyan square) are represented as a function of the time evolved for the Markov process. 
The succession of optimal severabilities at different time scales reveals the hierarchy of mesoscale structures containing a node of interest.
}
  \label{fig:art}
\end{figure*}

To introduce our method, we draw an analogy from energy landscapes \cite{wales2006energy}.
In particular, 
we consider the Markov landscape defined by the transition matrix of the standard random walk on a graph, where the nodes (or vertices) of the graph correspond to states and the landscape reflects the transition probabilities between them. 
Markov landscapes are analogous to energy landscapes although they lack a potential energy function pointing downwards to a minimum energy state.
Still, the notions of wells, barriers and roughness translate easily and helpfully to the language of time scale separation.
In this picture, a well is a group of states surrounded by high barriers (hence with a long escape time), whereas roughness inside the well is related to the mixing time (low roughness implies a fast mixing time).
An illustration of such a landscape can be found in Figure~\ref{fig:art}a, where we present a 3D representation of the luminosity landscapes of three paintings with very different characteristics; from a well compartmentalised painting by van Doesburg to a rough, featureless excerpt of Monet.
In this case, the barriers and roughness are obtained from differences in luminosity of adjacent pixels.
If a random walker (e.g. De Gennes' ant in a labyrinth \cite{de1976percolation}) is allowed to explore van Doesburg's luminosity landscape, the observed dynamics will reveal the presence of severable components in the state space; on the other hand, no such components would be expected in Monet's landscape. 

Mathematically, a subset of states of a system is defined to be a \textit{severable component} if it has both high barriers and low roughness, as extracted by the behaviour of the random walkers on the underlying landscape. As shown below in a precise sense (see~Section~\ref{sec:Sev_definition}),
 such a severable component can be understood as a \textit{mesoscale dynamical structure}, i.e., a set of states that behave coherently in the eyes of the external environment and which capture a relevant description of the system sitting between individual nodes and the global system.
To formalise these notions, we borrow the concepts of mixing and retention from Markov chain quasi-stationarity~\cite{darroch_quasi-stationary_1965}, as follows.
First, we introduce a measure of the mixing over a set of states $C$ by appealing to a random walker restricted to those states;
$C$ is poorly mixing over a timespan $t$ if the random walker's position at Markov times $0$ and $t$ are strongly correlated.
More precisely, we measure mixing by defining a quantity $0\le \mu(C,t)\le 1$ which measures the total variation distance between the probability distribution over $C$ at time $t$ and the \textit{quasi-stationary} distribution reached at long times, should the walkers remain in $C$ (Eq.~\eqref{eq:mu} in Methods). 
The mixing $\mu$ is thus inversely related to the roughness of the landscape over $C$,
since the exploration of $C$ is hindered by the roughness of the landscape. 
Secondly, we characterise the retention over the set $C$, which is directly related to the height of the barriers separating the set of states $C$ from the rest of the system, 
i.e., random walkers tend to stay within $C$ if it is hard for them to escape. 
This is quantified by $\rho(C,t)$, a number between $0$ and $1$ defined as the probability of a walker not escaping by time $t$ (Eq.~\eqref{eq:rho} in Methods).
Both $\rho$ and $\mu$ therefore range from 0 to 1, where the value of 1 corresponds to perfect retention or mixing, respectively. 

We now simply define the \textit{severability} of the set $C$ at time scale $t$ as
\begin{equation}
\label{eq:Severability}
\displaystyle \sigma(C,t) = \frac{\rho(C,t) + \mu(C,t)}{2}.
\end{equation}
Severability can be understood as a compound function that balances mixing and retention for a given set of states $C$ over the time scale $t$. If $C$ corresponds to a mesoscale dynamical structure, its severability will peak at some time $t_\textrm{max}$, below which the walkers are poorly mixed and beyond which retention is degraded.
In a connected network, the individual node and the entire graph will respectively have good severabilities for Markov time $0$ and $\infty$ for the trivial reason: at $t=0$, retention and mixing will be perfect for any individual node because nothing has diffused, and at $t=\infty$, the probabilities will have reached the ultimate stationary distribution, implying perfect mixing coupled with the always perfect retention of the entire graph.
At intermediate timescales, severable structures are of intermediate size, based on a combination of mixing and retention; on grid graphs, these optimally severable structures slowly expand with Markov time as higher times allow for mixing of larger diameter regions (Figure \ref{fig:art}).
Less uniform graphs have more interesting substructures; optimally severable structures remain so over a range of Markov times before jumping in size to another plateau.

This notion is illustrated in Figure~\ref{fig:art}b, where we show how the process of diffusion of information on a very simple model of a network, given by a hierarchical random graph with three levels of 16, 64 and 256 nodes, leads to severable components of the state space. 
As the time of diffusion increases, the random walkers gain sufficient probability to overcome the barriers of the landscape, and hence diffuse to larger portions of the network so that the optimal severable components grow from being single nodes at very short times, through each of the intermediate levels over different time scales, to the entire network at long times.

The components of an interconnected dynamical system with high severability have a precise mathematical meaning in terms of local time scale separation (see \textit{local time scale theorem} in Section~\ref{sec:theorem} in Methods). 
Briefly, the existence of a local time scale separation for a group of states in the dynamics of the random walker allows for a simplified model for the dynamical behaviour of the group of nodes $C$ when excited by an impulse, i.e. an arrival of probability mass into $C$. 
High retention and high mixing (implying high severability) at time $t_{\mathrm{max}}$  guarantees: (i)  the effect of $C$ on the rest of the system when given an impulse can be neglected altogether for time scales less than $t_{\mathrm{max}}$, and (ii) the subsystem $C$ can be accurately approximated to first order by a single state that aggregates all the states of $C$ for all times beyond $t_{\mathrm{max}}$.

In summary, the set $C$ can be thought of as a structure of intermediate size whose dynamical response to an impulse permits accurate simplified descriptions.
Our local time scale theorem (Section~\ref{sec:theorem}) is inspired by Simon and Ando's classic result for global time scale separation, yet it differs from it in that it seeks to find the conditions under which one can reproduce correctly the behaviour of a severable component at different time scales, independently of the rest of the system.
When the full interconnected system can be partitioned into components with \textit{comparable} time scales, we recover Simon and Ando's global theorem (see Supp Inf. B), demonstrating that our local time scale theorem generalizes their result.  

\subsection{Mesoscale components in power networks}
As a first application, we consider the synchronization dynamics of coupled nonlinear 
phase oscillators with Kuramoto-like sinusoidal coupling \cite{boccaletti2006complex}, which is found in areas as diverse as laser physics, biological synchrony of cells and animals, and power networks \cite{strogatz2000kuramoto}.
For our example, we will apply severability to a standard power network benchmark.
Power networks are composed of two types of nodes: generator buses, which deliver power, and load buses, which consume power.
The internal state of each node $i$ is described by a voltage, which oscillates with a frequency $\dot \theta_i$ around a nominal value (e.g. 50 or 60 Hz). The nonlinear dynamics of bus $i$ can be modelled as
\begin{equation}
M_i \ddot{\theta_i} + D_i \dot{\theta_i}= P_i - \sum_{j} A_{ij} \sin(\theta_i-\theta_j), 
\label{eq:smartgrids}
\end{equation}
where $M_i$ is an inertia (zero for some buses), $D_i$ is a damping coefficient, $P_i$ is the power being injected or withdrawn from the network at node $i$, and $A_{ij}$ indicates the strength of the (symmetric) interaction between $i$ and $j$~\cite{dorfler2013synchronization}.
Given sufficient coupling strength between the nodes, and depending on properties of the coupling matrix $A$, the network converges to a stationary state where all angles in the system oscillate at constant frequency $\omega$, keeping relatively small constant angle differences with respect to one another. 

\begin{figure*}[tbp]
\begin{center}
\includegraphics[width=0.9\textwidth]{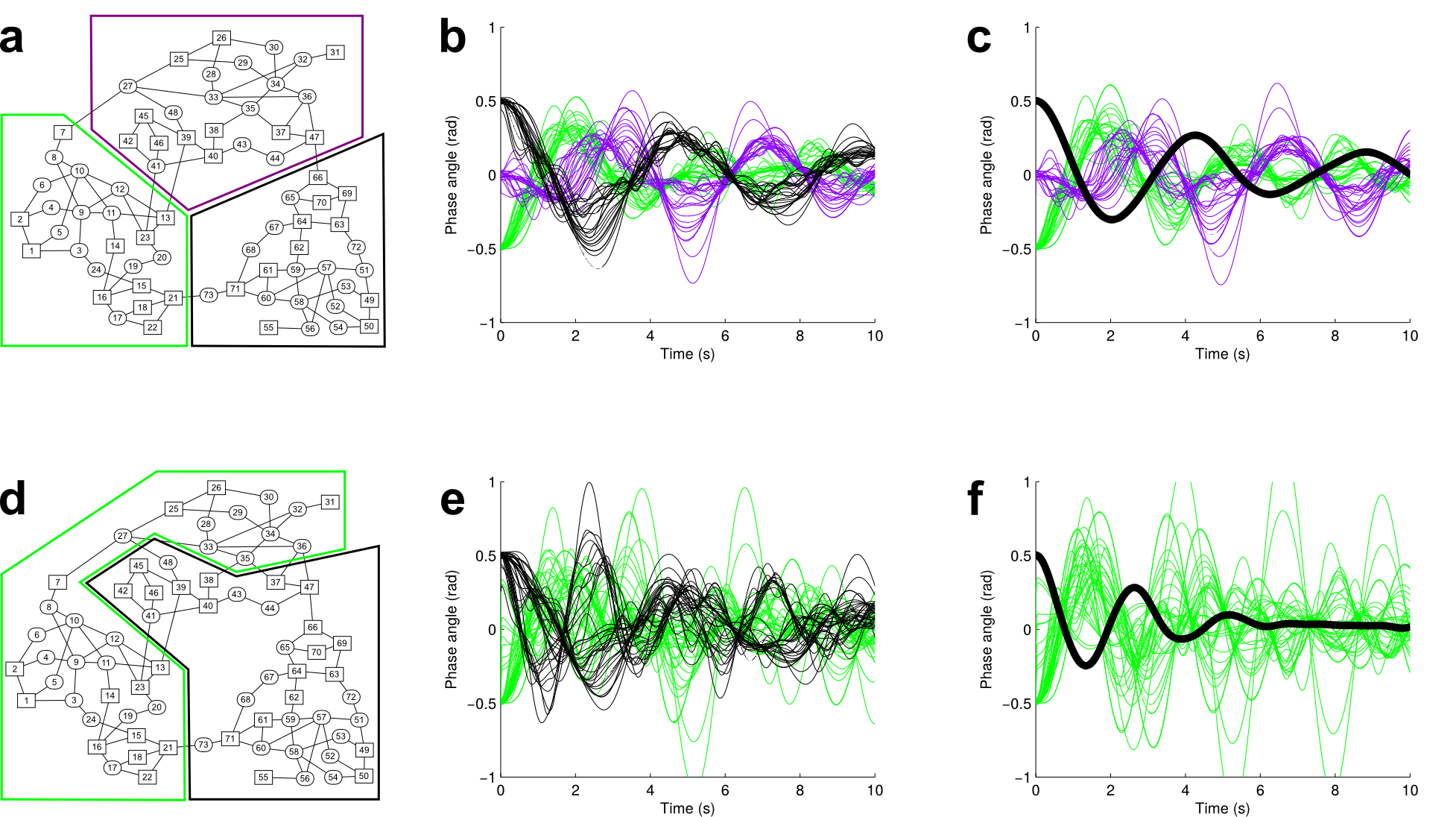}
    \caption{ Illustration of the three component RTS96 power network test system, which is composed of three copies of the RTS24 benchmark. Above, we have highlighted the three severable components as segments (at $t=64$, severability of 0.753, 0.753, and 0.807 for green, purple, and black respectively) (a), whereas below (d) we arbitrarily partitioned into two connected components (at $t=64$, severability of 0.611 and 0.646 for green and black respectively). (b,e) Full dynamics of the power network with starting phase angles chosen to match within a component. (c,f) Full dynamics of the power network using instead a collapsed state representing all of the black component. When the collapsed component is highly severable (top), the reduced representation matches the original system much better than when using arbitrary partitions (bottom).}
\label{fig:powernet}
\end{center}
\end{figure*}

Although this system is inherently nonlinear, severability is of use here.
In Figure~\ref{fig:powernet} we show the results of the application of our analysis to linearized discrete-time random walk dynamics based on node strengths $A_{ij}$, that is equivalent to the continuous-time nonlinear dynamics for small deviations around the synchronized state (see Supp.Inf.~\ref{sec:SIkuramoto} for details).
Our example is on a classic test case for power networks, the IEEE RTS96 test system, composed of three identical copies of the RTS24 test system interlinked with a few extra edges and one extra node.
Previous work has used time-scale based identification of \textit{global} partitions into slow-coherent areas based on global edge-counting or spectral methods~\cite{avramovic1980area,romeres2013novel}.
These global partitions correctly recover the expected components, but by nature require information about the entire network.
In contrast, severability recovers the expected components based solely on local information and provides a validation of the components in terms of their dynamical response.
More precisely, Fig. \ref{fig:powernet} shows that the fully nonlinear simulations of the model~(\ref{eq:smartgrids}) can be well represented by the aggregated angle variables within the components found with severability: the aggregation of angle variables within the `correct' components has little effect on the dynamics of the other variables of the system, whereas aggregation of an `incorrect' subgroup results in major discrepancies from the full dynamical evolution.
This result justifies the simplification of using random walk dynamics in severability, even for more complicated systems.

More generally, using random walks to model higher order dynamics provides
a general framework to capture central features of many other dynamics taking place on a network (Section \ref{sec:MCequiv}).

\subsection{On severable components and cliques}

\begin{figure}[tbp]
\begin{center}
\includegraphics[width=1\columnwidth]{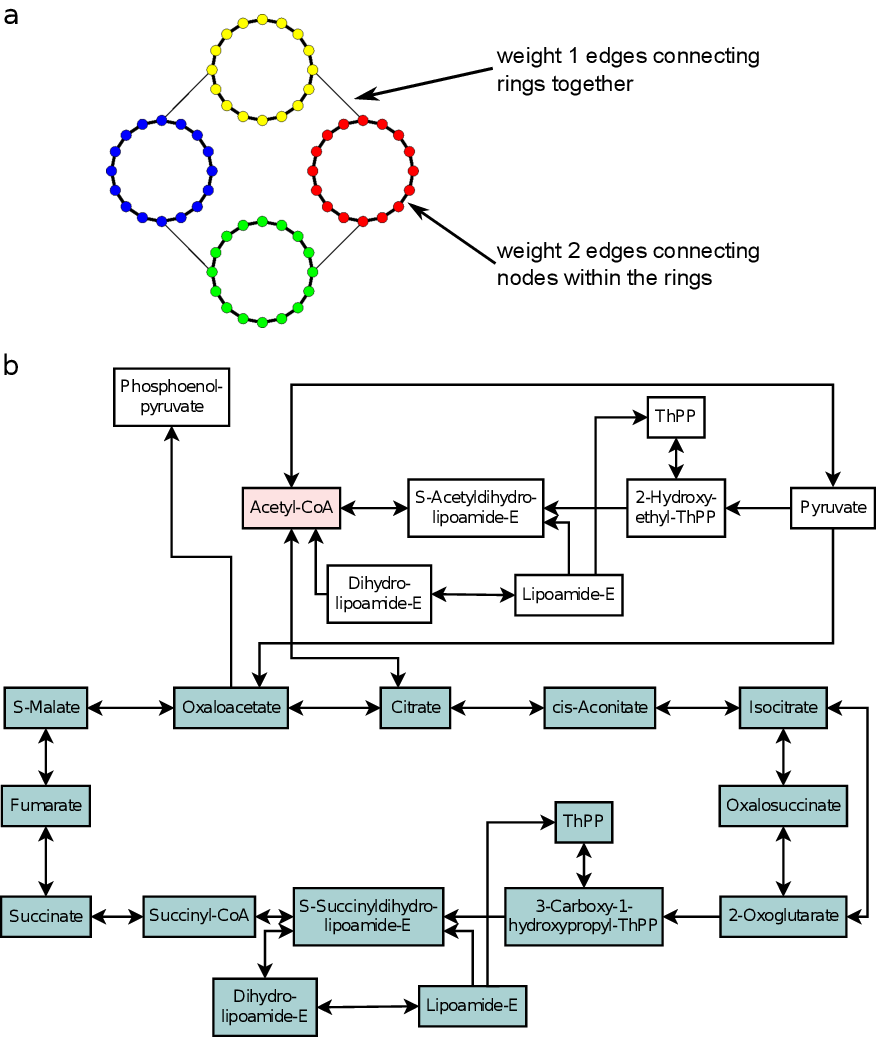}
\caption{(a)  Ring of rings. Heavy lines (within rings) correspond to undirected links with weight 2, while light lines between rings to links with weight 1. Severability is able to recover the seeded ring structure (at Markov times $3\le t<10$). %
(b) The citric acid cycle\cite{kanehisa_kegg:_2000}. The blue region is a stable component from Markov time $ 17 < t \le 21$ and adds acetyl-CoA from $21 < t \le 31$.}
\label{fig:rings}
\end{center}
\end{figure}

Given a network-centric view of severable components, it may come as no surprise that there are some similarities between network community detection and the discovery of severable components.
Network communities are groups of nodes with strong connectivity within the group and lower connectivity with the rest of the network.
Communities are often captured with local or global metrics that relate the number of edges crossing the boundaries of a community with the edges inside the communities, such as modularity~\cite{newman_finding_2004}, SBM  maximum likelihood~\cite{karrer2011stochastic}, OSLOM~\cite{lancichinetti2011finding} or conductance~\cite{shi2000normalized}, sometimes with random walk as a computational tool~\cite{andersen2007using,spielman2013local}.
Most of those criteria essentially capture the retention part, $\rho$, of severability. A few references \cite{kannan_clusterings:_2004,jeub2015think} also analyse the conductance internal to the cluster, which is a combinatorial criterion capturing essentially the mixing part $\mu$ of severability for $t=1$.
Although not all networks may have an easily measured dynamical interaction taking place on it (e.g. social networks), we can endow the graph with the standard random walk and apply severability to those dynamics.
Indeed, the standard random walk on a network can be used to approximate such things as opinion dynamics, information diffusion, and consensus problems, as the dynamics of the random walk is deeply related to the structure of the graph \cite{Delvenne13, rosvall_maps_2008,morarescu2009opinion,van2008graph,masuda2016random}.
As shown in Fig.~\ref{fig:art}b, the expectation is that such communities are detected as meaningful mesoscale components observed during information diffusion.
To validate this idea further, we have used the standard LFR synthetic benchmark network model for community detection, where networks are constructed as dense random Erd\H{o}s-R\'{e}nyi graphs interconnected by sparse random links at several levels of coarseness \cite{girvan_community_2002,lancichinetti_benchmark_2008}.
As shown in Supp.Inf.~\ref{sec:LFR}, our greedy algorithm for finding severable components recovers the communities with high fidelity, comparable or superior to other state-of-the-art methods~\cite{lancichinetti2009community}.
We remark that severable components are found from the local diffusive dynamics without global information from the graph. 

However, communities in social networks are often characterised by a clique-like structure, showing for instance a low diameter and high density of triangles \cite{palla_uncovering_2005}.
While clique-like structures emerge as particular cases of severable components, severability may detect long-range structures that are not akin to communities.
An example of this is shown in Fig.~\ref{fig:rings}a, where a ring-of-rings network is correctly revealed by severability.
Such non-clique like structures are present in other areas of application, including transportation networks, images, and protein structures~\cite{schaub2012markov}.
Another illustration is provided by biochemical networks, 
and one canonical metabolic pathway is the citric acid cycle.
When we analyze the citrate pathway schematic (map00020) in the KEGG database\cite{kanehisa_kegg:_2000} using severability,  
the search for high severability structures detects the Krebs citric acid cycle.
These structures do not fit the standard definition of communities, and indeed are not detected as such by most community detection algorithms \cite{schaub2012markov}.
However, as exemplified by the Krebs cycle, they are nonetheless dynamical structures of importance.
Thus, although severable components and network communities share some characteristics, they are different concepts built on different ideas, the former by coherency of dynamics, and the latter by the density of clique-like structure.

\subsection{Word association as a diffusion: overlaps and orphans}
\begin{figure*}[tbp]
\begin{center}
\includegraphics[width=1\textwidth]{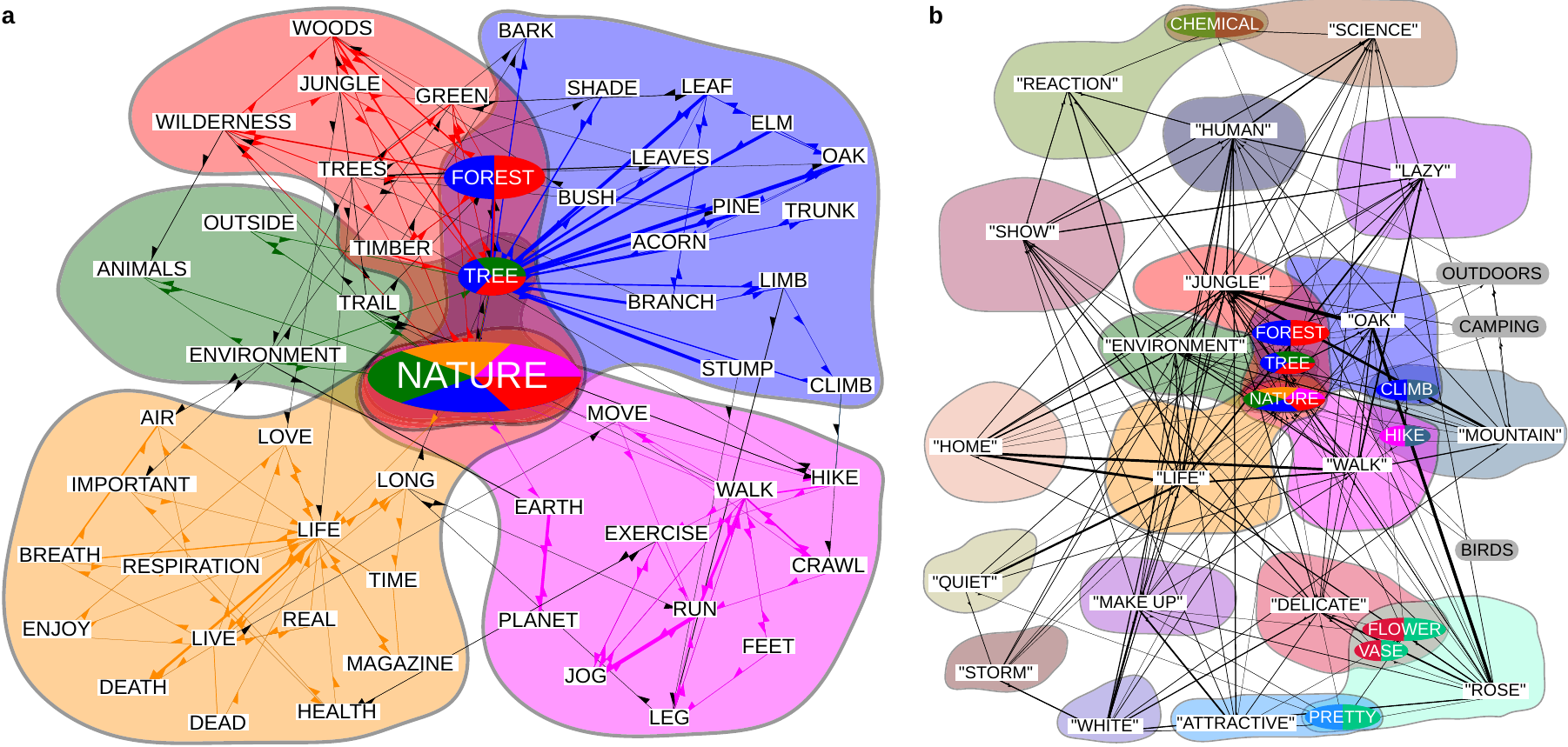}
\caption{(a) The five components that the word `nature' belongs to. Nodes and links are coloured by component identification; coloured ovals represent multiple component membership. (b) A broader view of the component landscape surrounding ``nature'', depicting also components connected to, but not containing, `nature', including three orphan nodes (see SI for details). Nodes belonging to just one of the components are combined into a single block labelled by the most central word of the component, while nodes belonging to more than one component are separately mentioned in the gray ovals. Note that in many cases, the words used to label the components possess multiple labels themselves. Communities were found by optimizing severability for Markov time $t=2$ and search size $S=50$.}
\label{fig:word_associate}
\end{center}
\end{figure*}

An important feature of severability as a means to analyzing interconnected sytems is that it allows the possibility of overlaps (a node can belong to more than one group) and orphans (a node can belong to no group, as every group that includes it has higher severability without it).
To illustrate these features,
we turn to a word association network (the University of South Florida Free Association Norms dataset~\cite{nelson_university_1998}), previously used to highlight the existence of overlapping network communities\cite{palla_uncovering_2005}.
To build this network, researchers presented words to participants, who were then asked for the first word that came to mind. Hence each node in the network corresponds to a word, and directed links between nodes are weighted according to the proportion of responses linking those two words. For example, when cued with `science', 21.4\% of participants wrote `biology'.

The very construction of the network is reminiscent of a random walk process representing the mental association based on similarity of meaning and contextual usage, thus making severability able to incorporate the weight and directionality of the network in a natural way.
As severability is a local method, it is not necessary to analyse the entire graph to find components.
Rather, by analysing increasing horizons on the network an expanding view of associated meanings presents itself from a particular vantage point.
Figure~\ref{fig:word_associate} shows the word `nature' and the components it belongs to (with maximum search size $S=50$ and Markov time $t=2$), as well as the components and orphan nodes to which `nature' is directly linked (see Supp.Inf.~\ref{sec:word_extended} for further details).
By permitting overlapping components, we are able to recover the different contexts and meanings associated with a single word.

\subsection{Locality in image segmentation:  zooming and cropping}
\begin{figure*}[tbp]
\begin{center}
\includegraphics[width=0.9\textwidth]{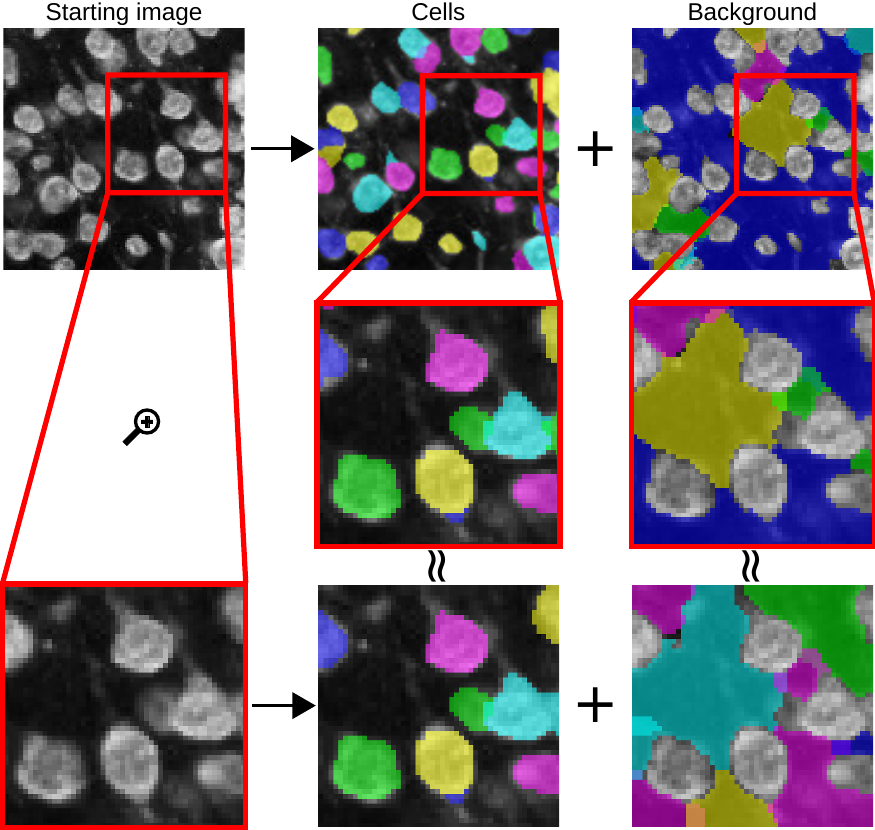}
\caption{Neocortical pyramidal neurons, stained with a fluorescent dye, with resolution reduced to $102 \times 102$ and converted to grayscale by luminosity. %
(Cyan) At Markov time $ t = 32$,  segments largely corresponding to cells were found (see S.I. for details). 
(Yellow) Furthermore, repeating the procedure with a cropped subregion of the image gives largely the same results, with some minor variations along the borders. This commutativity is a key feature of local methods.
Despite the fact that severability is not specifically designed for image analysis, the severable components found are of good quality.
}
\label{fig:images}   
\end{center}
\end{figure*}

In Fig.~\ref{fig:images}  we apply severability optimisation to 
the identification of stained neurons in a cell-fluorescence image  
in order to illustrate visually a central aspect of the method; namely, that it does not rely on global information in order to detect mesoscale components faithfully.
Below we show that the results are similar whether the algorithm is run on only some part of the image, or on its entirety.

Image segmentation divides images into subsets of adjacent pixels of similar color or luminosity, and is particularly used for medical and biological imaging. Some of the existing segmentation methods are based on a nominal diffusion dynamics taking place on the lattice graph 
of pixels~\cite{grady2006random,schaub2012markov}.
In this view, a segment can be seen as a particular case of a severable component, as already  suggested in our initial view of the paintings in Fig.~\ref{fig:art}. 
To carry out our analysis, we have followed a classic protocol to generate a lattice graph from the image by assigning an edge between pixels weighted by a function of the difference in luminosity and distance (up to a cutoff) \cite{shi2000normalized, browet_community_2011, wu1993optimal} (for details see Supp.Inf.~\ref{sec:image_proc}).
The severable subsets are self-consistent in that they are found robustly from diffusions starting from any of the members of the set; as these are strongly severable subsets, they tend to be found regardless of which member of the set is used as a starting point (unlike the word association communities of the last section).
Both the cells and patches of the background are found as severable components.
Furthermore, because severability does not depend on global information, the results do not change significantly when the algorithm is run only on a smaller section of the image:
only segments that lie on the edges of the image are affected by cropping.
This feature is of potential application to evolving networks, as communities are stable against perturbations and do not need to be recomputed fully when new nodes are added outside of a local neighbourhood.

We note that while other completely local methods\cite{jeub2015think} exhibit a similar commutativity, `mostly' local methods like OSLOM (Order Statistics Local Optimization Method) do not \cite{lancichinetti2011finding}.
Though OSLOM is based on local order statistics, when partitioning, it takes into account some global information, making it noncommutative.
In Appendix \ref{sec:smallworlds}, we show that OSLOM behaves differently on a ring of small-world networks vs a single small-world.

\section{Discussion}

Real-life dynamics emerging from the interaction of many elementary nodes can sometimes be seen as the interconnection of mesoscale dynamical structures, whose evolution over a time scale of interest can be represented as a single aggregated state interacting with its surroundings.
We have introduced a measure for the detection of such severable components which can be well approximated by their aggregated variables.
The formal theory, a generalization of Simon and Ando's classic theory to local time scale separation is illustrated on the particular case of Markov chains, which are representative of a larger class of dynamics, including consensus and synchronization. 
Severable components, which can coexist at several sizes and time scales, overlap and leave orphan nodes.
This dynamical concept is connected to other more particular notions encountered in several classes of systems, including basins (energy landscapes), slow-coherent areas (power networks), segments (image processing), communities (social networks analysis), and rings (biochemical networks) which can be understood as structures with a locally coherent dynamics.
On the other hand, other kinds of meso-structures in complex networks (e.g., block models, or roles in ecological systems \cite{beguerisse2014interest,cooper2010role,cooper2011role})
are global in nature, and do not fall under the condition of locality required in this paper.
Hence, while locality is an advantage for large systems, by definition truly global characteristics cannot be thus discovered.
However, we have shown in this paper that many classes of structures are in fact locally defined, 
demonstrating the applicability of severability.

Engineering disciplines traditionally operate by plugging together smaller components, usually seen as black boxes with simple external behaviour regardless of their internal complexity, 
in order to generate complex systems with controlled behaviour. One may argue that many natural systems are built similarly. In this perspective, we aim here to reverse this process: 
although complex systems are often too large to analyse in their entirety, our approach here is to try and find if there exist suitable intermediate dynamical components which provide a proper understanding and representation of the complex global dynamics.
In this sense, severability serves the role of a local coarse-graining mechanism for the dynamics as observed from a given subset of states in the system. 
Appealing to the coexistence of local time scales in Markov processes as a means to reveal severable components establishes mathematical connections between diffusion processes and model reduction, linking in a precise sense good mixing and retention in a subsystem to its accurate approximation through coarse-graining while preserving the Markov property.

As Big Data continues to proliferate, severability provides a first step towards the definition of new methods able to tackle the huge wealth of data being collected in all areas of science, technology and social life, much of which comes with a naturally endowed dynamics. 
Undoubtedly, more challenges lie on the road ahead, such as in the treatment of more sophisticated node dynamics, for example when the dynamics are strongly nonlinear or non-Markovian \cite{rosvall2014memory,delvenne2015diffusion}. 
Yet the importance of dynamics as a key to characterising networks will undoubtedly persist.
Ultimately, we hope that the framework of severable components (code available at \url{https://github.com/yunwilliamyu/severability}) provides not only a specific solution to recovering mesoscale structures when the dynamics are roughly Markovian, but also a meaningful and practical starting point for more sophisticated methods capable of tackling these more difficult problems.

\clearpage

\section{Methods}
\subsection{Formal definition of Severability}
\label{sec:Sev_definition}
The definition of severability uses concepts of graph theory and Markov chains.
A graph $G$ is a set $V$ of $n$ nodes (or vertices, or states in the Markov chain terminology) together with another set $E$ of links (or edges) between vertices. 
We assume that every node has at least one outgoing edge, and that all edges are labelled with a positive weight. The weighted, directed graph, is encoded as an adjacency matrix $A$, where $A_{ij}$ is the weight of the edge going from $i$ to $j$. The (weighted) out-degree of node $i$ is the sum of weights of edges leaving $i$. The out-degrees can be compiled in the vector $A \mathbf{1}$, where $\bm{1}$ is the $n \times 1$ vector of ones and $D = \textrm{diag}(A\bm{1})$ is the diagonal matrix of out-degrees.

On a given graph $G$, we define a random process in discrete time. A random walker starts from a node $i$ at time $0$ and jumps at time $t=1$ to any out-neighbour $j$ with probability 
$A_{ij}/d_i$, proportional to the edge weight.  Successive jumps at $t=1, 2, 3, 4,\ldots$ define a Markov chain, or random walk, on the graph. The probability of presence of the random walker evolves as 
\begin{equation} \label{eq:xxDAP}
x(t+1) = x(t) \, D^{-1} A \equiv x(t) \, P,
\end{equation}
where $x(t)$ is the $1 \times n$ normalised probability vector and $P$ is the transition matrix, the rows of which are non-negative and sum to one.
Provided that the graph is strongly connected and aperiodic (i.e. there is no integer $k>1$ such that all cycles comprise an exact multiple of $k$ edges), any initial probability distribution converges to a unique stationary distribution, which is a solution of the fixed-point equation $x(\infty)=x(\infty) P$.

Given a connected subset $C \subset V$ with $k$ nodes, 
let $Q$ be the submatrix of $P$ corresponding to the nodes in $C$. 
Then we define the \emph{retention} of the subset $C$ over time $t$,
$\rho(C,t)$, as the probability for a random walker starting with a uniform probability distribution in $C$ not to have escaped by time $t$:
\begin{equation}
\label{eq:rho}
\rho(C,t) = \frac{1}{k} \left( \mathbf{1}^\mathrm{T} Q^{t} \mathbf{1} \right).
\end{equation}

To define mixing,  let $q_i^{(t)}$ be the $i$th row of the matrix $Q^t$. Note that because $C$ is connected, $q_i^{(t)} \neq \mathbf{0}$. Thus the normalised row of $Q$, $q_i^{(t)}/ q_i^{(t)} \mathbf{1}$  is the probability distribution at time $t$ for a random walker starting from node $i$, conditional upon the walker remaining in $C$ between 0 and t. We can then define the internal \emph{mixing} $ \mu (C,t)$ as 
\begin{equation}
\label{eq:mu}
\displaystyle \mu (C,t) = 1- \frac{1}{k} \sum_{i=1}^k \left \| \bar{q} - \frac{q_{i}^{(t)}}{q_i^{(t)} \bm{1}} \right \|_{TV},
\end{equation}
where $\bar{q}$ is the arithmetic mean over the unit-normalised rows of $Q^t$, and we have used the fact that the total variation distance norm is given by 
\begin{equation}
    \displaystyle  \left\| \bm{v} \right\|_{TV} = \frac{1}{2} \sum_i |v_i| .
\end{equation}
The internal mixing term $\mu$ approaches 1 as the probability distribution of a random walker starting somewhere uniformly random within the community approaches the quasistationary distribution on that subset of nodes \cite{darroch_quasi-stationary_1965}.

Both $\rho$ and $\mu$ are defined to range from 0 to 1, where the  value of 1 corresponds to perfect retention or mixing, respectively. We define the  \textit{severability} as compound function of both retention and mixing:
\begin{equation}
\sigma(C,t) = \frac{\rho(C,t) + \mu(C,t)}{2},
\end{equation}
which can be understood as the quality of the subset $C$ to be considered as a separate dynamical mesostructure over time $t$.
Severability has an intrinsic resolution parameter $t$, corresponding to the Markov horizon; as $t$ increases, the random walker will diffuse to larger parts of the graph, as reflected by the iterations of the submatrix $Q$. Note that, from the above definitions, $\sigma(C,t)$ depends only upon the out-links from nodes within $C$; hence it is a purely local function.

The particular form assumed for retention, mixing and severability is justified by the mathematical properties stated in \ref{sec:theorem}, and proved in the Supp.Inf.

\subsection{Local time scale separation theorem}
\label{sec:theorem}

\subsubsection{Background: Simon and Ando's global time scale separation theorem}
\label{sec:globaltheorem}

In 1961, Simon and Ando established a time scale separation theorem, both for general linear systems and  Markov chains in particular~\cite{SimonAndo61}, which we present now.

Given a Markov chain 
\begin{equation} \label{eq:xxP}
x(t+1)=x(t) \, P,\end{equation}   let us split the nodes in two sets 
$$x(t)=\begin{pmatrix} x_1(t) & x_2(t)\end{pmatrix},$$ 
with a corresponding partition of $P$ 
\begin{equation}
P=\begin{pmatrix} 
P_{11} & P_{12} \\
P_{21} & P_{22} 
\end{pmatrix}.
\label{eq:PP}
\end{equation}
Fix an arbitrary $\epsilon>0$, which will serve as a requested standard of approximation. Assume that $P$ is close to a perfectly decoupled transition matrix
\begin{equation}
\tilde{P}=\begin{pmatrix} 
\tilde{P}_{11} & 0 \\
0 & \tilde{P}_{22} 
\end{pmatrix}.
\label{eq:PPtilde}
\end{equation}

Simon and Ando proved that there is a small enough $\delta (\epsilon, \tilde{P}_{11}, \tilde{P}_{22}) > 0 $ 
 and a time $T(\epsilon, \tilde{P}_{11}, \tilde{P}_{22})$  such that if $\|P-\tilde{P}\| \leq \delta$, 
 then two kinds of approximations are valid for the trajectories of $x(t)$:
\begin{itemize}
\item On the one hand, for all times $t \leq T$, the decoupled approximation 
\begin{equation}\label{eq:approxshort}
  x_{dec}(t)=\begin{pmatrix} x_{dec,1}(t) & x_{dec,2}(t) \end{pmatrix} = \begin{pmatrix} x_1(0) P_{11}^t  &  x_2(0) P_{22}^t  \end{pmatrix}
\end{equation} is within $\epsilon$  in norm from the actual solution $x(t)$:
\begin{equation}\label{eq:errorapproxshort}
 \| x(t) - x_{dec}(t) \| < \epsilon, \quad t<T
\end{equation}
\item On the other hand, and more importantly, for all times $t$ and in particular for all $t>T$, the aggregated probabilities 
$$x_{tot}(t)=\begin{pmatrix} x_{1, \textrm{tot}}(t) &  x_{2, \textrm{tot}}(t) \end{pmatrix} =
\begin{pmatrix} x_{1}(t) \mathbf{1} &  x_{2}(t) \mathbf{1} \end{pmatrix}$$ 
are within $\epsilon$ in norm from the approximation
\begin{equation}\label{eq:approxlong} 
x_\textrm{tot,approx}(t)= \begin{pmatrix} x_{1, \textrm{tot,approx}}(0) & x_{2, \textrm{tot,approx}}(0) \end{pmatrix} \begin{pmatrix} 
\lambda_1 & \delta_{12} \\
\delta_{21} & \lambda_2 
\end{pmatrix}^{t},
\end{equation}
for some real values $\lambda_1, \lambda_2, \, \delta_{12},\delta_{21}$.
\item Moreover, for times $t >T$, $x_i(t)$ can be reconstructed as $x_{i,\textrm{tot}}(t) v_i$ with an error bounded by $\epsilon$, for some $v_i$.
\end{itemize}


Which norms are chosen in the statement above is irrelevant, as all norms of vectors or matrices of a given size are equivalent up to a factor, making the statement true for any choice of them.
For simplicity we stated here the two block case in a Markov chain dynamics, although the theory holds for general linear systems split in an arbitrary number of blocks.
It is important to notice that the required $\delta$ depends not only on the given $\epsilon$ but also potentially on \textit{all} the entries of the diagonal blocks $\tilde{P}_{ii}$, as is apparent for example in our own proof of Simon-Ando's theorem (Supp.Inf.~\ref{sec:globaltheorem}). 
The theorem can therefore only be applied globally, with full knowledge of the dynamics. It is desirable to decouple this global condition into the local conditions to be satisfied by each diagonal block $P_{ii}$ to satisfy the required accuracy $\epsilon$, and severability offers one practical way to achieve this, as shown below.

\subsubsection{Statement of the local time scale theorem}
\label{sec:theorem_statement}

Following the same notation as above, consider the first block $P_{11}$ (denoted $Q$ in the main text and Methods A), which describes the set of states $C$ with severability $\sigma(t)$. 
The local dynamics of $x_1(t)$ is described by the open dynamical system 
\begin{equation}
\begin{aligned}
x_1(t+1)&=x_1(t) P_{11} + u_1(t) \\
y_1(t)&=x_1\jcdbis{(t)} P_{12},
\label{eq:open}
\end{aligned}
\end{equation}
where $u_1$, defined for all $t \geq 0$, is the input into the subsystem $C$, i.e., the in-flow of probability from the environment, 
and $y_1$ is the output of the subsystem, by which it influences the environment by an outflow of probability.
By the environment we mean the rest of the state space, described by  $x_2$, itself governed 
by an open system equation of the same kind as Eq.\eqref{eq:open}. The global dynamics on 
$\begin{pmatrix} x_1(t) & x_2 (t)\end{pmatrix}$ can be understood as the feedback interconnection of the two systems, related by the equations: $u_1(t)=y_2(t), \,  u_2(t)=y_1(t)$.

Equation~\eqref{eq:open} describes a relationship between the input sequence $u_1(t)$ and the output sequence $y_1(t)$. An alternative way to describe an open system in linear response theory is by its  \emph{impulse response}. 
In our notation, the impulse response can be written as $$g(t)=P_{11}^{t-1} P_{12}.$$                       for all $t\geq 1$, and zero for $t \leq 0$. 
The impulse response characterizes fully the input-output relationship in that the output generated by any input sequence $u_1(t)$ is obtained by the convolution product $y_1=u_1 * g$ (defined as $y_1(t)=u_1(t)g(0) + u_1(t-1)g(1) + u_1(t-2)g(2) + \ldots$ for all $t\geq 0$), assuming zero probability in $C$ initially, $x_1(0)=0$. A non-zero initial state can be incorporated by adding an artifical input $u(-1)$, setting a state $x_1(0)=u(-1)$.
To approximate the behaviour of $y_1$ described by  Eq.~\eqref{eq:open}, we need to approximate the function $g(t)$ by another impulse response $h(t)$ between the same input and output spaces measured with a given norm. A common metric used in the open systems literature is the one-norm 
\begin{equation}
\|g-h\|_1=\sum_{t\geq 0} \|g(t)-h(t)\|,
\label{eq:L2}
\end{equation}
whenever it is defined.
Of course, given matrices $g(t), h(t)$, we can choose any matrix norm for $\|g(t)-h(t)\|$,  as they all relate within a constant factor only dependent on the dimension of the matrix. If the approximation is only meant to be valid on time interval $[t_1,t_2]$, then we can restrict the sum in Eq.~\eqref{eq:L2} to 
$t \in [t_1,t_2]$, denoted $\|g-h\|_{1,[t_1,t_2]}$. 
An error in the impulse response committed in replacing $g$ by $h$ will result in an error in the output $y_1$ in the following way, as one can show from elementary algebra: $$\sup_{0 \leq t \leq \overline{t}} \|(u_1 * g)(t) - (u_1 * h)(t)\| \leq   \sup_{t} u_1(t) \cdot  \|g-h\|_{1,[0,\overline{t}]},$$
where $\overline{t}$ can be infinity.

Our local time scale separation theorem makes two statements, regarding the approximability of the impulse response of the nodes $C$, before and after an arbitrary chosen time $T$. 
The first one at short times follows directly from the high retention implied by a high severability at time $T$, whereas the second one at long times requires a more careful analysis. The theorems are proved in Supp. Inf.~\ref{sec:proof_local}.

\begin{LocalTheo}[Short times]
The system represented by Eq.~\eqref{eq:open} can be approximated until time $T$ by the trivial response $y(t)=0$, with accuracy $\|g-0\|_{1,[0,T]}=\mathcal{O}(1-\sigma(T))$.
\end{LocalTheo}

In other words, the influence of the system on its environment can be neglected altogether over short time scales.

\begin{LocalTheo}[Long times]
The system described by Eq.~\eqref{eq:open} can be approximated by a one-state system of the following form
\begin{equation} \label{eq:onedim}
\begin{aligned}
x_C(t+1)&=\lambda \, x_C(t) + u_1(t) b\\
y_1(t)&= x_C(t) \, d,
\end{aligned}
\end{equation}
where $\lambda$ is the dominant eigenvalue of $P_{11}$ and $b,d$ are appropriate vectors, whose corresponding impulse response is $h(t)=b \lambda^t d$. Vectors $b,d$ are found from the dominant eigenvectors $P_{11}b=\lambda b$, $vP_{11}=\lambda v$, and $d=vP_{12}$, normalised so that $vb=1=v\bm{1}$.  The approximation is valid for all times---including obviously  $t>T$---and the error summed over all times is $\|g-h\|_1 = \mathcal{O}(1-\sigma(T))$.
\end{LocalTheo}
For any given input signal $u_1(t)$ bounded by $\|u_1(t)\| \leq K$ for all $t \geq -1$, the exact model described by Eq.~\eqref{eq:open} and the one-dimensional model given by Eq.~\eqref{eq:onedim} deliver outputs whose difference is at all times bounded by $\mathcal{O}(1-\sigma(T))K$.

The constants contained inside $\mathcal{O}(.)$ in these statements may depend on the dimension of $x_1$ (number of nodes in $C$), but neither on the specific entries of $P$ nor on $T$. 
In view of these statements, the best time scale separation is given 
by $T=t_{\mathrm{max}}$, at which severability peaks, and the error of the resulting 
approximations is $1-\sigma(t_{\mathrm{max}})$.  

Assuming now that the global network is split into two or several blocks, one may combine the different local approximations and obtain the following version of classic Simon-Ando theorem: given a global dynamics given by Eqs \eqref{eq:xxP} and \eqref{eq:PP}, suppose that we find a common time $T$ at which 
both $1-\sigma(T,P_{11}) \leq \delta$ and $1-\sigma(T,P_{22}) \leq \delta$, then the short-term and long-term dynamics can be approximated as Eqs \eqref{eq:approxshort} and \eqref{eq:approxlong} with error bounded by $\epsilon =\mathcal{O}(\delta)$, where the hidden constant only depends on the total number of nodes (Supp. Inf.~\ref{sec:proof_Ando_new}). The generalisation to more than two components is straightforward.  This version highlights the role of severability of each component and the need to find a common global time scale $T$ (possibly suboptimal for each component separately) where each component simultaneously reaches a high severability, for a global time scale separation to emerge. 

See Supp. Inf.~\ref{sec:4by4example} for a toy example of comparative application of the global and local time scale separation theorems.

\subsection{Computational aspects of Severability optimisation}
\label{sec:computational}
We apply a semi-greedy search algorithm to find the optimal component $C$ for a starting node $n_0$, at a chosen Markov time $t$ and setting a search size $S$ (see Appendix \ref{sec:optimisation} in the Supp. Inf. for a detailed flowchart). 

Briefly, the algorithm proceeds as follows.
Without loss of generality, define $\sigma(C)= \sigma(Q,t)$.
Initially, only $n_0 \in C$.  
Aggregate nodes greedily, except let every third step be a Kernighan-Lin switch of a single node on the boundary of $C$ to maximise $\sigma(C)$ \cite{kernighan_efficient_1970}. 
After the initial semi-greedy optimisation, the intermediate component $\mathfrak{C}$ that has maximal severability is fine-tuned using Kernighan-Lin switches to find a local maximum.
If $n_0$ is in the resulting component, the algorithm stops; otherwise, start over with a different neighbour of the starting node.
If all neighbours of $n_0$ have been attempted without success, declare $n_0$ an orphan.
For the word association network, every neighbour of ``nature'' was attempted for the first step, giving the overlapping communities.

A detailed description of other computational aspects of the implementation are discussed in Supp. Inf.~\ref{sec:implement}.

\subsection{Markov chain equivalence of dynamical systems}
\label{sec:MCequiv}

Markov chains, or random walks, are characterized by a dynamics of the form 
$x(t+1)=x(t) P$, where $P$ is any nonnegative square matrix with all rows summing to one.
To every such dynamics we can associate a dual \emph{consensus} dynamics $y(t+1)= P y(t)$ acting on the column vector $y(t)$, the entries of which are positions, or opinions, of agents, which converge to one another until convergence to the same value if and only if the corresponding random walk converges to a unique stationary distribution. 

Positive linear systems are common in economics, biology, chemistry, where variables naturally take nonnegative values. Such systems are characterized by an evolution  $y(t+1)=Py(t)$, or $x(t+1)=x(t) P$, where $P$ is only required to be nonnegative. 
Under the same connectivity conditions on the network
underlying $P$, we know that there is a unique dominant eigenvalue $\lambda$ and a corresponding left and right eigenvectors, $u=\lambda^{-1}uP$ and $v=\lambda^{-1}Pv$ respectively, all of which are positive by virtue of the Perron-Frobenius theorem. 

This property allows a normalization that transforms the dynamics into a consensus, or random walk, dynamics. The new matrix is $\tilde{P}=\lambda^{-1} D_v^{-1} P D_v$, where $D_v$ is the diagonal matrix associated to $v$. It is readily observed that $\tilde{P}$ is a valid transition matrix, and is equivalent to $P$ except for a global scaling $\lambda^{-1}$ and a change of variable on every node. In particular, it has the same eigenvectors and acts on the same underlying network as $P$. This transformation has an elegant information-theoretic interpretation as the random walk with maximal entropy rate (if $P$ is a zero-one matrix) or a free energy (if the nonnegative entries are interpreted as exponential energy barriers along the edges) 
\cite{Ruelle1978,delvenne2011centrality}.

Markov chains are also defined in continuous time, following an equation of the form
$\dot{x}(t)=x(t) L$, where the continuous-time transition matrix has nonpositive diagonal, nonnegative off-diagonal terms, and zero-sum rows. One also has continuous-time consensus, and any positive continuous-time linear system, characterized by a matrix $L$ with nonpositive diagonal and nonnegative off-diagonal terms, can be similarly normalized to a continuous-time Markov chain, which can be sampled to a discrete-time Markov chain.

Some non-linear systems can be linearized around a fixed point. Classic theorems such as Hartman-Grobman's ensure that the nonlinear and linearized systems are equivalent up to a change of variables in a neighbourhood of the fixed point. Kuramoto oscillators, and power networks dynamics, linearize to consensus dynamics.

\begin{acknowledgments}
The authors would like to thank Antoine Delmotte, Michael Schaub, Arnaud Browet, Florian D\"orfler and Renaud Lambiotte for code and/or discussions.
Neuron fluorescence imagery is courtesy of Simon Schultz and Marie-Therese Vasilache.
Y.W.Y. was partially supported by an Imperial Marshall Scholarship during the early years of this work.
J.-C. D. is partly supported by the Flagship European Research Area Network (FLAG-ERA) Joint Transnational Call “FuturICT 2.0”.
This work is partially supported by the Engineering and Physical Sciences Research Council of the United Kingdom.

    \textbf{Author Contributions}

    M.B. and S.Y. conceived the project and guided the research. J.-C.D. developed the theoretical analyses. Y.W.Y. implemented and designed the experimental analyses. The manuscript was jointly written by all authors.

\end{acknowledgments}

\bibliographystyle{apsrmp}
\bibliography{short}

\clearpage

\appendix
\onecolumngrid
\begin{center}
\Large{Supplementary Information}
\end{center}
\bigskip

\section{Proof of the local time scale separation theorem}
\label{sec:proof_local}
\begin{proof}
On the short time scale, the validity of the approximation is given by $T \|g(t)\|$ where $(t)=P_{11}^{t-1}t P_{12}$ for $t\geq 1$ (and $0$ at $t=0$).
 
We notice that $\sum_{t=1^T} P_{11}^{t-1}P_{12}$ expresses the probabilities of escape within time $T$. In fact retention $\rho(T)$ as introduced in Eq.~\eqref{eq:rho} can be expressed as
\begin{equation}
\begin{aligned}
1-\rho(T)&=\frac{1}{k} \bm{1}^\mathrm{T} \sum_{t=0}^{T} P_{11}^t P_{12} \bm{1}\\
         &=\sum_{t=0}^{T+1} \|g(t)\| \\
         &\geq  \sum_{t=0}^{T} \|g(t)\|
\end{aligned}
\end{equation}
for the choice of $k$-by-$k$ matrix norm $\|A\|=\sum_{ij} |A_{ij}|/k$, given that all entries of $g(t)$ are nonnegative.
The short time scale local theorem results from $1-\rho(T) \leq 2(1-\sigma(T))$ which follows directly from the definition of severability.

Equation~\eqref{eq:onedim} generates an impulse response $h(t)= b \lambda^{t-1} d$ for $t \geq 1$. 
The difference $\Delta(t) = \|P_{11}^{t-1} P_{12} -b \lambda^{t-1} d \|$, decays to zero exponentially provided that $P_{11}$ has dominant eigenvalue $\lambda$, eigenvectors $v= \lambda^{-1} v P_{11}$  (normalised so that the entries of row vector $v$ sum to one), $b= \lambda^{-1} P_{11} b$ (normalised so that $vb=1$) and $\jcdbis{d}= v P_{12}$. Indeed this guarantees that $P_{11}^t$ behaves as $b \lambda^t v$ for high $t$.

 If $P_{11}$ were perfectly stochastic, then $P_{11} \bm{1}=\bm{1}$ and we would have $b=\bm{1}$ and $\lambda=1$. As $P_{11}$ is almost stochastic, we expect that $1-\lambda$ and $\bm{1}-b$ are in $\mathcal{O}(1- \sigma(t))$ for all $t$, which we can prove indeed in the following way.  

It is well known from Perron-Frobenius theory that the dominant eigenvalue of a matrix with positive entries sits between the minimum and maximum row sum.Therefore $k(1-\rho(t)) \leq \lambda^t < \lambda < 1$, therefore $1-\lambda \leq \mathcal{O}(1-\sigma(t))$.
To evaluate $b$, let us call $\tilde{P}_{11}(t)$ the row-normalized matrix derived from $P_{11}^t$, where every row is scaled so as to sum to one. Then the distance (in any norm) between any two rows of $\tilde{P}_{11}(t)$ is in $\mathcal{O}(1-\mu(t))$, by the definition of internal mixing in Eq.~\eqref{eq:mu}. The distance between $P^t_{11}$ and $\tilde{P}_{11}(t)$ on the other hand is in $\mathcal{O}(1-\rho(t))$, by definition of the retention. 
Therefore the distance between any two rows of $P^t_{11}$ is in $\mathcal{O}(1-\sigma(t))$, thus $P_{11}^t= \bm{1} v_0 + \mathcal{O}(1-\sigma(t))$ for some positive vector $v_0$. 
Premultiplying this equality by $v$, we get $\lambda^t v=v_0 + \mathcal{O}(1-\sigma(t))$, thus $v=v_0 + \mathcal{O}(1-\sigma(t))$. Postmultiplying instead by $b$ gets $b= \bm{1} + \mathcal{O}(1-\sigma(t))$, as required.

Consider the remainder $R= P_{11}-b \lambda v$, thus $R^t= P_{11}^t-b \lambda^t v$ from spectral decomposition properties.  We find from the above that $R^t=\tilde{P}_{11}(t)-\bm{1}v_0 + \mathcal{O}(1-\sigma(t))= \mathcal{O}(1-\sigma(t))$.

Choosing the matrix norm  $\|A\|=\sup_i \sum_j |A_{ij}|$, which happens to be submultiplicative ($\|AB\| \leq \|A\|\|B\|$ for all $A,B$),
and using the identity $\sum_{t \geq 0} z^{t} = (1- z)^{-1}$, applied here to $z=R^T$ (for which the identity is valid because all eigenvalues of $R^T$ have absolute value $\leq\lambda<1$), we deduce a bound for the error on the impulse response: 
\begin{align*}
\|g-h\|_1 &= \sum_{t \geq 1} \Delta(t)\\
&= \sum_{t \geq 1} \| R^{t-1} P_{12} \| \\
&\leq (1-\|R^T\|)^{-1} ( \|P_{12}\|+\|RP_{12}\|+\cdots+\|R^{T-1}P_{12}\| )\\
&=  (1-\|R^T\|)^{-1} \mathcal{O}(\|(I+P_{11}+ \cdots + P_{11}^{T-1})P_{12} \|) \\
&=\mathcal{O}(1) \mathcal{O}(1-\sigma(T)),
\end{align*}
    using also $R^t=\mathcal{O}(||P_{11}^t||)$ and $\|\sum_i A^{(i)}\| \leq \sum_i \|A^{(i)}\| \leq k \|\sum_i A^{(i)}\|$ for any family of $k$-by-$k$ nonnegative matrices $A^{(i)}$.

    We obtain the same result $\|g-\tilde{h}\|_1=\mathcal{O}(1-\sigma(T))$ for $\tilde{h} = \tilde{b} \lambda^{t-1} d$, with $\tilde{b}=\bm{1}$.
This is because $\|\tilde{h}-h\|_1 = \mathcal{O}(1-\sigma(T))$, easily obtained from the fact $\tilde{b}-b=\mathcal{O}(1-\sigma(T))$.  This approximation has the nice property of preserving the flow of probability: it describes the behaviour of a single super-node that aggregates all the input probability flows, expelling a small fraction $1-\lambda$ of stored probability mass at every step to the nodes in the rest of the system, with weights given by $d$.
\end{proof}

Therefore different approximations rule the short-term (where a large retention matters) and long-term behaviours (where fast mixing also matters).

The proof highlights that the theorem is robust with respect to the choice of norms in the definition of severability and in the statement of the theorems, as it changes the hidden constants in a way that only depends on $k$, the number of nodes. The specific definition chosen for severability in this article is motivated by simplicity, convenient computation and good practical results.

\onecolumngrid

\section{From local to global: a proof of Simon-Ando's theorem}
\label{sec:proof_Ando_new}

We now provide a proof of Simon and Ando's global time scale theorem, stated in terms of severability of the components.
Assume that a partition of the network into two components reveals a common time scale $T$ at which each severability is higher than $1-\epsilon$. 
In the short run, every component can ignore the other and evolve separately, with a resulting error of order $\mathcal{O}(\epsilon)$. Let us turn to the long run case.


For times $t\geq T$, one may write $x_1(t)=x_1(t-T) P_{11}^T  + \mathcal{O}(\epsilon)$   (as $\sum_{k=t-T}^t u_1(k)$ the probability mass leaking from component $x_2$, is in $\mathcal{O}(\epsilon)$). From high mixing, all rows  of $P_{11}^T$ are $\epsilon$-close to one another, and  $\epsilon$-close to a multiple of  the dominant eigenvector $v_1$ (quasi-stationary distribution). The same holds for $x_2$, $\epsilon$-close to a multiple of $v_2$. The full state trajectory $x(t)=(x_1(t) \,\,\, x_2(t))$ thus remains $\epsilon$-close to a trajectory of the form $(\alpha(t) v_1   \,\,\, \beta(t) v_2)$, and therefore it is enough to know the two-dimensional trajectory $(\alpha(t), \beta(t))$  (in fact one-dimensional in the set of probability measures because subject to the constraint $\alpha(t)+\beta(t)=1$) to reconstruct approximately $x(t)$. This means that $S$, the image of the set of all probability measures under the map $P^T$ is invariant under $P$,  has diameter $1-\mathcal{O}(\epsilon)$ in the direction $\{(\alpha v_1\,\,\, (1-\alpha)v_2) | \alpha \in \mathbb{R} ) \}$, and is `thin' in that every point of $S$ is $\epsilon$-away from that direction. 

Now consider the two-dimensional dominant eigenspace of $P$, generated by the dominant left eigenvector (stationary distribution) $w^{(1)}$ of eigenvalue $1$ and the second left eigenvector $w^{(2)}$ (normalised to unit norm) of eigenvalue $1-\mathcal{O}(\epsilon)$. The intersection of that space in the space of probability measures is one-dimensional, of the form $S_0=\{w^{(1)}+ \gamma w^{(2)}|w^{(1)}+ \gamma w^{(2)}\geq 0\}$. On this eigen-set, the dynamics takes the simple, exact form $x(t)= w^{(1)}+ \lambda^t \gamma w^{(2)}$. Given that $S_0 \subseteq S$, we know that every point $x= (x_1 \,\,\,  x_2) \in S_0$ is $\mathcal{O}(\epsilon)$-approximated by the projection on $\textrm{Proj}(x)=( x_1\bm{1}  v_1 \,\,\,  x_2 \bm{1} v_2)$. 
Therefore, the aggregated dynamics obtained in replacing $w^{(1)}$ and $w^{(1)}$ by their approximations in terms of $v_1$ and $v_2$ induces a one-dimensional aggregated dynamics on the direction $(\alpha v_1\,\,\, (1-\alpha)v_2)$ where $x_1$ and $x_2$ are replaced by their aggregation 
$x_1\bm{1}  v_1$ and $x_2 \bm{1} v_2$, and the projected dynamics is given by $\textrm{Proj}(x(t))=\textrm{Proj}(w^{(1)}) + \lambda^t \gamma \textrm{Proj}(w^{(2)})$.

 The trajectory initiated by a point $x \in S_0$, and the trajectory generated by the aggregated by its projection  $\textrm{Proj}(x)$ with this projected dynamics, remain  $\mathcal{O}(\epsilon)$-close at all times.

On the other hand, any point $x$ in $S$ is  $\mathcal{O}(\epsilon)$-close to a point $x_0$ in $S_0$, and those two points remain $\mathcal{O}(\epsilon)$-close when both are iterated by $P$, as $P$ contracts the 1-distance (or total variation distance ;  the induced 1-norm of $P$ is 1).

 Now we can conclude. The trajectory initiated by any point in $S$ (iterated by the exact dynamics $P$) remain $\mathcal{O}(\epsilon)$-close at all times from some trajectory in the eigen-set $S_0$, which itself remains  $\mathcal{O}(\epsilon)$-close at all times from the projected, aggregated dynamics on the direction $(\alpha v_1\,\,\, (1-\alpha)v_2)$. Therefore any trajectory in $S$ generated by the actual dynamics $S$ is $\mathcal{O}(\epsilon)$-close at all times from the one-dimensional dynamics on the aggregated quantities. 

A closer look would show that the projected dynamics taken from the approximation given by the Local Time Scale separation theorem on each block separately, is not strictly identical to the aggregated dynamics presented here, but the trajectories generated by the two one-dimensional dynamics are  $\mathcal{O}(\epsilon)$-close at all times again. 

In the above, all hidden constants in the $\mathcal{O}(.)$ notation are dependent on the specific norms used to measure distances, thus dependent on the  number of nodes in each block, but on nothing else.

This completes the proof of Simon and Ando's global time scale theorem, as given in Methods (see Section \ref{sec:globaltheorem}), since arbitrarily small perturbations from a fixed, block-diagonal transition matrix $\tilde{P}$ lead to arbitrarily high severability for arbitrarily large intervals of time. 

The global nature of the theorem reveals itself in the fact that it needs simultaneously at time $T$, a high mixing  and a high retention in every component, thus shedding light on the conditions required for global time-scale separation to hold.

See next Appendix for a simple example showing that $\delta(\epsilon,\tilde{P}_{11}, \tilde{P}_{22})$ and $T(\epsilon,\tilde{P}_{11}, \tilde{P}_{22})$ described in the text in the classic statement of Simon-Ando's theorem indeed depend on the global information $(\tilde{P}_{11}, \tilde{P}_{22})$. 

\clearpage
\section{Global vs Local time scale separation theorems: an example}
\label{sec:4by4example}

We apply our version of Simon-Ando's theorem (formulated in terms of severabilities) to a toy example of four nodes separated into two blocks, or mesoscale components. We then modify the example so that Simon-Ando's  global theorem does not apply any more, but our local theorem still applies.

 Consider 
 \begin{equation}
P=\begin{pmatrix} 
1-\eta_1 - \delta & \eta_1     & \delta    & 0 \\
\eta_1            & 1 - \eta_1 & 0        & 0\\
0                &  0          & 1-\eta_2  & \eta_2\\
0                &  \delta     &  \eta_2  &  1-\eta_2-\delta  \\
\end{pmatrix}, 
\end{equation}
which is $\delta$-close to the block-diagonal matrix   
\begin{equation}
\tilde{P}=\begin{pmatrix} 
1-\eta_1 & \eta_1 & 0        & 0  \\
\eta_1 & 1-\eta_1  & 0        & 0 \\
0      &     0   & 1-\eta_2 & \eta_2 \\
0      &    0    &  \eta_2   & 1-\eta_2
\end{pmatrix}. 
\end{equation} 
Let us compare the trajectories generated by the two initial conditions $(1 \,\, 0\,\, 0\, 0)$  and
$(0 \,\, 1\,\, 0\, 0)$, which both lead to the same aggregation (probability 1 in the first block). 

If $\eta_1 \ll \delta \ll 1$ and $\eta_2 \ll \delta \ll 1$, then it is clear that their trajectories will remain very different, even at the aggregated level, for a long time, as  at times of the order $1/\delta$, the first trajectory will be concentrated mostly on the second block (and so will the aggregated trajectory), while the second trajectory will stay confined in the first block. 

If $\eta_1 \ll \delta \ll \eta_2 \ll 1$ then at time at  times of the order $1/\delta$ the first trajectory will be equally split between the two blocks, while the the first trajectory will be again confined in the first block.

Thus if we want to reach a given accuracy in Simon-Ando's theorem, for instance $\epsilon=0.1$,  we need to take $\delta$  of the order of $\min(\eta_1, \eta_2)$, which shows the global dependency of $\delta$ on the `internal details' of both blocks. The transition between the short time regime and the long time regimes  occurs at time $T=\mathcal{O}(1/\delta)$.   

In our language, the severability of each block $i=1,2$  will be high (close to one) for times $t$ of between  $\mathcal{O}(1/\eta_i)$ and $\mathcal{O}(1/\delta)$ (if $\delta \ll \eta_i$ indeed, otherwise the severability remains low at all times). We see indeed that these intervals will start overlapping at time $\mathcal{O}(1/\delta)$. We can therefore apply our version of Simon-Ando's theorem, as we have   simultaneous high severability in each block, for some time $t$. 

This also shows the intrinsically asymptotic nature of Simon-Ando's original theorem: as $\delta$ is decreased, the peak of severability for each block extends into a plateau stretching until $1/\delta$, eventually forcing overlap of plateaus for small enough $\delta$. 

If we consider a slightly more complicated example:
 \begin{equation}
P=\begin{pmatrix} 
1-\eta_1 - \delta_1 & \eta_1     & \delta_1    & 0 \\
\eta_1            & 1 - \eta_1 & 0        & 0\\
0                &  0          & 1-\eta_2  & \eta_2\\
0                &  \delta_2     &  \eta_2  &  1-\eta_2-\delta_2  \\
\end{pmatrix}, 
\end{equation}
with  $\delta_2 \ll \eta_2 \ll \delta_1 \ll \eta_1$ then Simon-Ando's theorem cannot be formally applied, because it assumes a \emph{fixed} block-diagonal structure, and an arbitrarily small perturbation of it.
We find the same conclusion in the language of severability. We see that the severability of each block $i=1,2$ peaks in the interval of times between   $\mathcal{O}(1/\eta_i)$ and $\mathcal{O}(1/\delta_i)$. As these intervals do not coincide, we indeed cannot apply our version of Simon-Ando's theorem. 

Our local time scale theorem is nevertheless applicable to each block separately, and allows us to identify them as  \emph{mesoscale components} reaching high severability at \emph{different} time scales. This shows that the local time scale theorem is of wider applicability and is a more relevant tool to identify components with dynamical coherence in a complex, heterogenous dynamical system.

%


\clearpage

\section{Computational aspects of Severability}
\label{sec:implement}
\subsection{Severability optimization flowchart}
\label{sec:optimisation}
\begin{figure*}[h]
  \begin{center}
    \includegraphics[width=.95\textwidth]{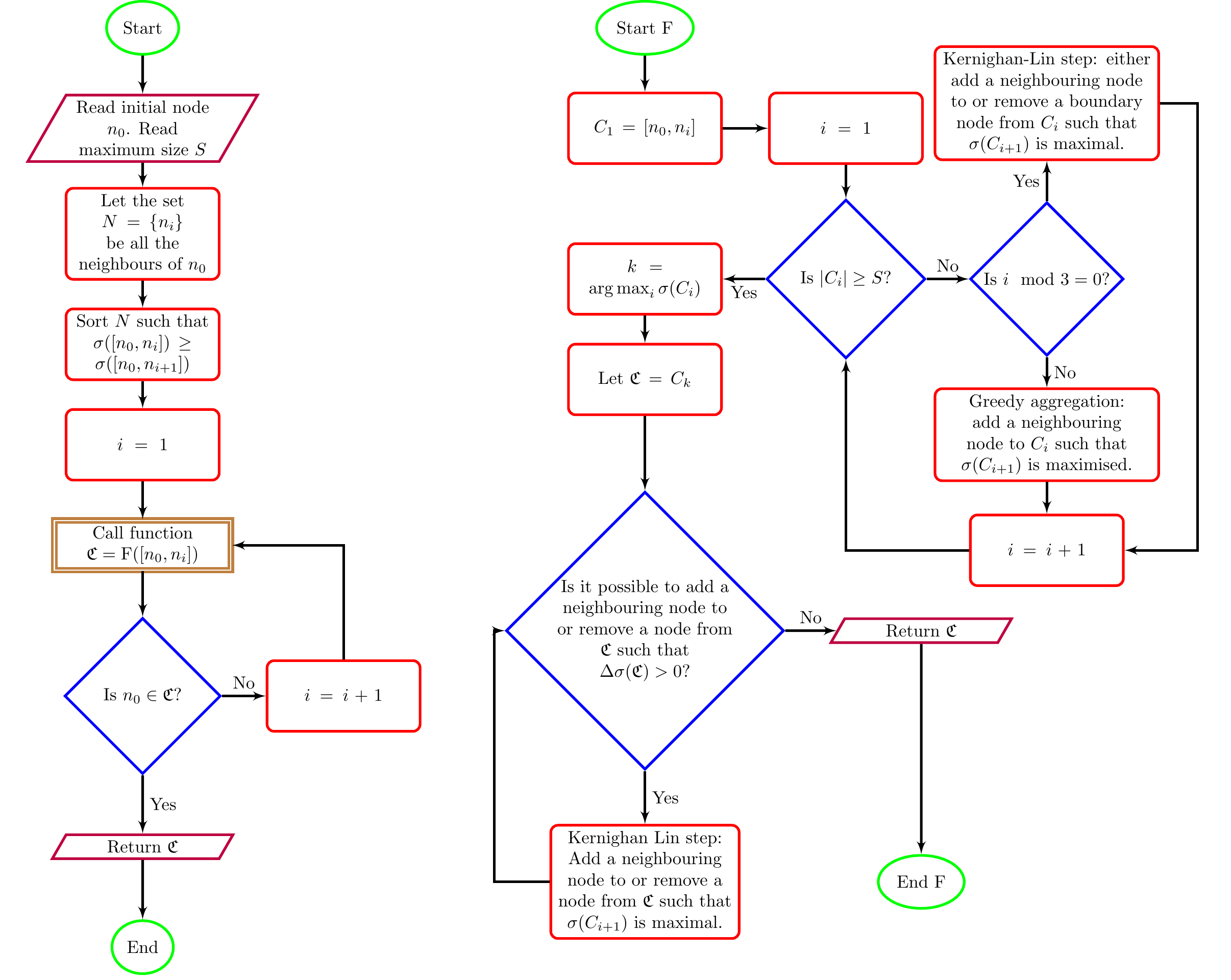}
    \caption{Flowchart of the optimisation procedure to find the most severable component to which a node $n_0$ belongs. For clarity,  the Markov time $t$ is assumed to be constant in this diagram.}
    \label{fig:optimisation}
  \end{center}
\end{figure*}

\subsection{Computational Complexity}
Let $n$ be the number of nodes in a graph.
The severability of a component $C$ of size $k$ for a Markov time $t$ can be computed in $P(k,t) = O(k^3 \log_2 t)$ time, where the cubic term comes from schoolbook matrix multiplication.
Computation of mixing and retention given $Q^t$ are both O($k^2$) operations, so the total cost is dominated by matrix exponentiation.

The cost can be reduced using fast matrix multiplications techniques; for instance, using Strassen's method, the total cost would only be $O(k^{2.807} \log_2 t)$.
Alternatively, for large $t$, matrix diagonalisation can be first employed, which makes the $t$ term negligible, giving a $O(k^3)$ solution.

However, finding good components is more involved than simply computing the severability of a single set of nodes.
The cost of the component optimisation algorithm described in Appendix \ref{sec:optimisation} is more difficult to characterise, as it depends strongly on the number of nodes neighbouring the putative component throughout the procedure.
In pathological cases, the cost is $O( n S \cdot P(S,t)) =O(n S^4 \log_2 t) $, where $S$ is the maximum number of nodes permitted in the component, and $n$ is the size of the graph.
Luckily, this upper bound only occurs in complete graphs, and so is of little relevance as most real networks are far sparser.
However, by specifying the maximum component size $S$, one can choose the maximal computational resources one wants to spend trying to find a component. 

Potential optimizations include using a random walk to highlight likely candidate neighbours; for instance, by choosing only the $l$ nodes that a random walker uniformly distributed in $C$ would most likely walk to in the next step, or for removal of nodes, the $l$ nodes in $C$ that have the least density of probability.
Such an algorithm would only cost $O( S\, P(S,t))$, a significant improvement.

More subtly, the computational cost of the matrix powers might also be reduced, by taking advantage of the fact that $Q(C*)$ for each of the neighbouring components is effectively a rank-2 perturbation of $Q(C)$.
Furthermore, as briefly mentioned in the discussion, severability is only one way of quantifying the mixing and retention of random walkers. Other, alternate, methods may be found that are quicker.

\subsection{Benchmarking against community detection methods}
\paragraph*{Optimal component cover.}
To compare against benchmarks with overlapping components, it is necessary to generate a list of components to cover the network.
Simply taking the optimal components of each node is suboptimal, because then there are many duplicate components in the list.
Instead, we chose the following naive method:
\begin{enumerate}
    \item Let $components = {}$ be the set of components; let $covered$ be the set of nodes that have been assigned to at least one component.
    \item Choose a node $x$ that is more connected to unassgined nodes than to nodes in $covered$. If no such node exists, end.
    \item Find the optimal component $C(x)$ for $x$, and add $C(x)$ to $components$ and the nodes in $C(x)$ to $covered$.
    \item Repeat from step 2.
\end{enumerate}

\paragraph*{Partitioning.}
To compare severability with partitioning methods, it is necessary to turn the optimal component cover into a partition.
To do so, first order the components of the cover arbitrarily.
Where a node appears in multiple components, always choose the first component it appears in.
This procedure is obviously dependent upon the ordering of the components; however, in networks with well-defined partition structure, this method works sufficiently well, as demonstrated in the LFR benchmark.

\paragraph*{Choice of Markov Time.}
For hierarchical networks, Markov time serves as a useful resolution parameter, allowing for severability to pick out optimal component structure at different levels.
However, existing metrics \cite{danon_comparing_2005,lancichinetti_detecting_2009} require the selection of a single time $t$.
For partitions, this can be done by choosing a Markov time to minimise the number of singleton and overlapping vertices, but other $t$ could be chosen.

\paragraph*{Quantifying similarity of partitions.}
To compare partitions across different methods, normalised mutual information \cite{danon_comparing_2005} has been employed.
To compare component covers, a generalisation of normalised mutual information that allows for overlapping nodes has been used \cite{lancichinetti_detecting_2009}.
We refer to the generalised variant as simply ``normalised mutual information'', without loss of precision as only the generalised variant can be used in the benchmarks with overlapping components.

\clearpage

\twocolumngrid
\section{Linearization and dicretization of a network of Kuramoto Oscillators} \label{sec:SIkuramoto}
For power networks, in a number of situations of practical relevance~\cite{dorfler2011topological,dorfler2012synchronization}, e.g. when operating in the regime where frequencies $\dot{\theta}_i$ have almost synchronized, the term $M_i \ddot{\theta_i}$  can be reasonably neglected and one may linearize around the steady state trajectory to obtain 
\begin{equation}
 \Delta  \dot \theta_i = \sum_j D_i^{-1} A_{ij} \Delta \theta_j = \sum_j L_{ij} \Delta \theta_j  ,
\label{eq:consensus}
\end{equation}
where $A_{ii}$ is defined as $-\sum_{j \neq i} A_{ij}$.
The matrix $L$ is called the Laplacian of the network, as it plays the same role in graphs as the Laplace operator in continuous space.
It is important to note that this equation also fully characterizes the consensus model of opinion dynamics~\cite{ROS-JAF-RMM:07}, the heat equation, and random walkers diffusing through the network in continuous time \cite{chung1997spectral}; to wit, the $\theta_i$ represent, respectively, converging opinions, equalizing temperatures, or the expected fraction of walkers on node $i$ at any given time. 

In order to build a discrete-time random walk to which our framework can be directly applied, we choose a timestep $\delta = 0.02 \cdot 2^{-\eta}$, where $\eta$ is the smallest natural number such that a modified adjacency matrix $A' = 0.02 \cdot 2^{-\eta} L + I$ is strictly positive. 
We then measure the severability of random walk dynamics on the graph defined by the modified adjacency matrix $A'$.

\section{Variants of the LFR benchmark}
\label{sec:LFR}
\subsection{Unweighted, undirected, non-overlapping LFR networks}
We analyse a class of networks in which components are extremely unevenly sized, a situation in which many popular partitioning methods perform suboptimally. 
These multi-scale networks are randomly constructed such that both degree and component size distributions follow power laws, with exponents $\gamma$ and $\beta$, respectively. 
Additional parameters include the total number of nodes $N$, the average degree $\langle k \rangle$, the maximum degree $k_{max}$, and the intrinsic parameter $\mu$---not to be confused with the mixing $\mu(C,t)$ which is part of severability. 
The fraction of links from a node to other nodes within the same component is given by $1-\mu$ \cite{lancichinetti_benchmark_2008}.
Graph generation parameters were chosen at values typical of real networks: $\gamma = 2$, $\beta = 2$, $N = 1000$, $\langle k \rangle = 15$, and $k_{max}=50$ \cite{lancichinetti_benchmark_2008}.
Severability optimisation was performed with a maximum search size $S=50$, and partitions were generated from the component cover.

\begin{figure}[tbp]
  \begin{center}
\includegraphics[width=1\columnwidth]{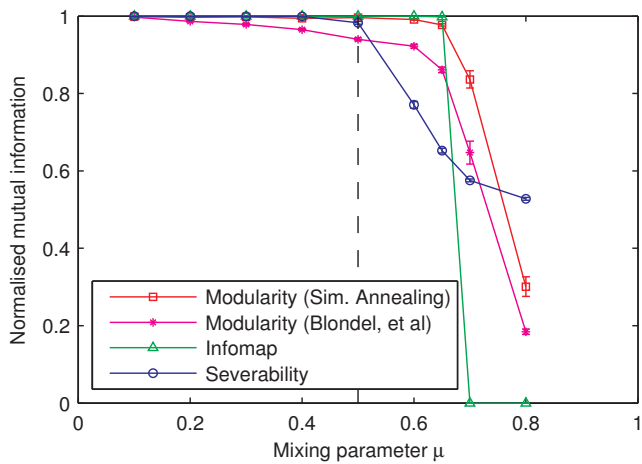}
    \caption{Comparison of severability with modularity and infomap the LFR benchmarks with exponents $\gamma = 2$, $\beta =2$, average degree $\langle k \rangle = 20$, and maximum component size of 50. Severability optimisation was performed with maximum search size of 50, and Markov time $t = 3$ (a value determined as a result of minimising the number of orphan nodes and overlapping nodes). Modularity was optimised for using both simulated annealing\cite{good_performance_2010}, which is extremely slow, but gives good results, and a faster heuristic by Blondel, et al \cite{blondel_fast_2008}. Each point is an average over ten random realisations.}
    \label{fig:fortunato}
  \end{center}
\end{figure}

As can be seen in Figure~\ref{fig:fortunato}, severability performs well, always finding the natural component structure up to until around $\mu = 0.5$, when components are no longer defined in a strong sense~\cite{radicchi_defining_2004}.
That severability begins failing at $\mu = 0.5$ is as expected and consistent with its definition, since at that point random walkers are as likely to escape during each step as to remain within any of the pre-seeded components.
Recalling the definition, a component is defined as severable precisely when random walkers tend to stay and mix within it.
Even so, the results are comparable to that of Infomap and modularity optimisation using simulated annealing, which have been found to be amongst the most successful methods for this benchmark \cite{lancichinetti2009community}.

\subsection{Unweighted, undirected, overlapping LFR networks}
\begin{figure}[tbp]
  \begin{center}
\includegraphics[width=1\columnwidth]{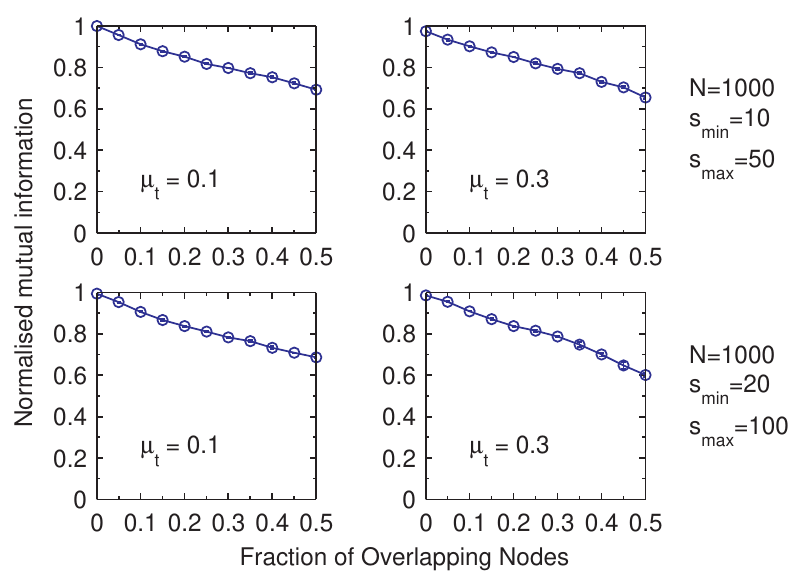}
    \caption{Severability at Markov time $t = 4$, with an unweighted, undirected, overlapping variant of the LF Benchmark\cite{lancichinetti_benchmarks_2009}. The networks have 1000 nodes; the other parameters are $\tau_1 = 2$, $\tau_2 = 1$, $\langle k \rangle = 20$, $k_{max} = 50$. Each point is an average over five random realisations.}
    \label{fig:binary_overlap}
  \end{center}
\end{figure}

Further extensions to the LFR benchmark were implemented to allow for components to overlap \cite{lancichinetti_benchmarks_2009}. 
In Figure~\ref{fig:binary_overlap}, we compare the component covers from severability to the pre-seeded components.
For the optimisation, the maximum search size $S = 50, 100$ was used for the upper and lower panels, respectively.
The parameters chosen were identical to those used for the evaluation of k-clique percolation\cite{palla_uncovering_2005} in figure 6 of Ref.~\cite{lancichinetti2009community}. Comparison with those results shows that severability performs comparably for the smaller component sizes, but significantly better for larger components.

\subsection{Weighted, directed, overlapping LFR networks}
\begin{figure}[tbp]
  \begin{center}
\includegraphics[width=1\columnwidth]{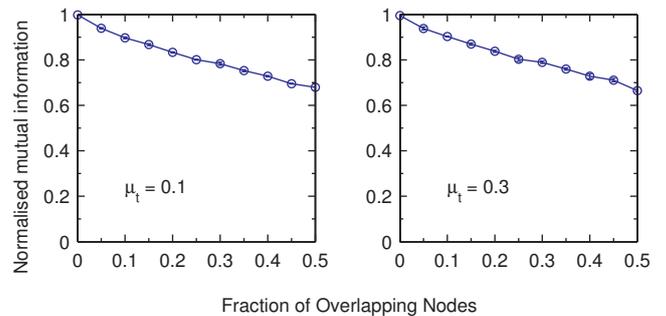}
    \caption{Severability at Markov time $t = 4$, with a weighted, directed, overlapping variant of the LF Benchmark\cite{lancichinetti_benchmarks_2009}. The networks have 1000 nodes; the other parameters are $\tau_1 = 2$, $\tau_2 = 1$, $\mu_w = 0.2$, $\langle k \rangle = 20$, $k_{max} = 50$, $s_{min} = 20$, $s_{max} = 100$. Each point is an average over five random realisations.}
    \label{fig:weighted_directed}
  \end{center}
\end{figure}

Severability also loses no accuracy when direction and weight are added to the benchmark~\cite{lancichinetti_benchmarks_2009} (as seen in Figure~\ref{fig:weighted_directed}). This is expected, since the Markov chain formulation naturally includes both. For the optimisation shown, the maximum search size $S = 100$.

\twocolumngrid

\twocolumngrid

\twocolumngrid

\section{Image processing}
\label{sec:image_proc}
The image in Fig. \ref{fig:images} of the main text  was pre-processed by reducing the image resolution to a more convenient size and converting to a network using standard methods. Briefly, we connect only adjacent pixels (using the maximum metric) with link weight 
$ w =  \exp \left[ - (\Delta I)^2/\sigma_I^2 \right],$ where $\Delta I$ is the difference in luminosity and $\sigma_I $ is an adjustable parameter controlling the exponential weight decay. 
Here we used $\sigma_I = 20$. Severability was optimized with Markov time $ t = 32$, and maximum size $s=200$.

In a post-processing step, segments with mixing $\mu > 0.9$ and retention $\rho < 0.1$ were removed as outliers, because at high Markov times they correspond to nearly disconnected components. If a component $C_1$ was completed embedded in $C_2$ ($C_1 \subset C_2$), we keep only the one with higher severability.
Communities were then inductively merged if they overlapped by more than 20 pixels until no more merges were possible. 
Merging is generally relevant when a feature of the network is much larger than the maximum search size; in this case the optimisation method gives overlapping patches of the background, which can then be pieced together.
The segments were ordered by average luminosity, and the darker patches were assigned to the background.

\section{Ring-of-rings}
\begin{figure}[tbp]
\begin{center}
\includegraphics[width=1\columnwidth]{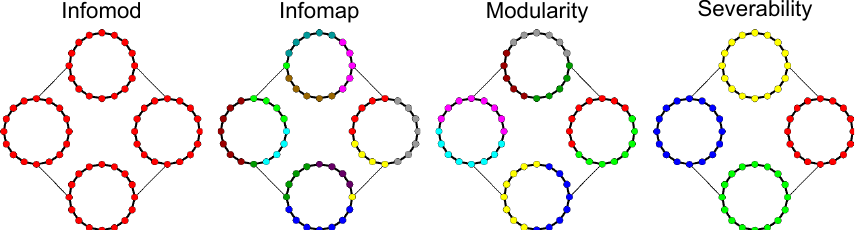}
\caption{ Ring of rings. As in Figure \ref{fig:rings}, heavy lines (within rings) correspond to undirected links with weight 2, while light lines between rings to links with weight 1. Severability is able to recover the seeded ring structure (at Markov times $3\le t<10$). Infomod returns the entire network, whereas Infomap and Modularity each return only arcs from the rings.}
\label{fig:rings_supp}
\end{center}
\end{figure}

We also examined the results of running several other popular graph partitioning methods on the ring-of-rings network shown in Figure \ref{fig:rings}. Infomod, Infomap, and Modularity were all unable to recover the ring structure of the graph (Figure \ref{fig:rings_supp}).

\section{Square lattice}

As a negative control, It is instructive to consider a network in which there is clearly no structure. For that, we chose a regular 2-D square lattice with each node connected to all 8 neighbours (including diagonal links). 
We visualise this using a uniformly coloured discrete image, in which each pixel is connected to all of the adjacent pixels with links of equal strength.
As can be seen in the figure below, after accounting for symmetry considerations, all components found are transients, which is the expected result.

Additionally, these images strongly suggest a relationship between severability optimisation and diffusion. This is of course quite closely related both to the dependance of severability on random walk dynamics and to the optimisation procedure outlined in Appendix \ref{sec:optimisation}. Along these lines, the optimisation procedure we outlined can be thought of as a modified random walk in which previously explored states are immediately accessible to the random walker, but probability barriers in the ``energy landscape'' are magnified.

\begin{figure}[tbp]
\begin{center}
  \includegraphics[width=1\columnwidth]{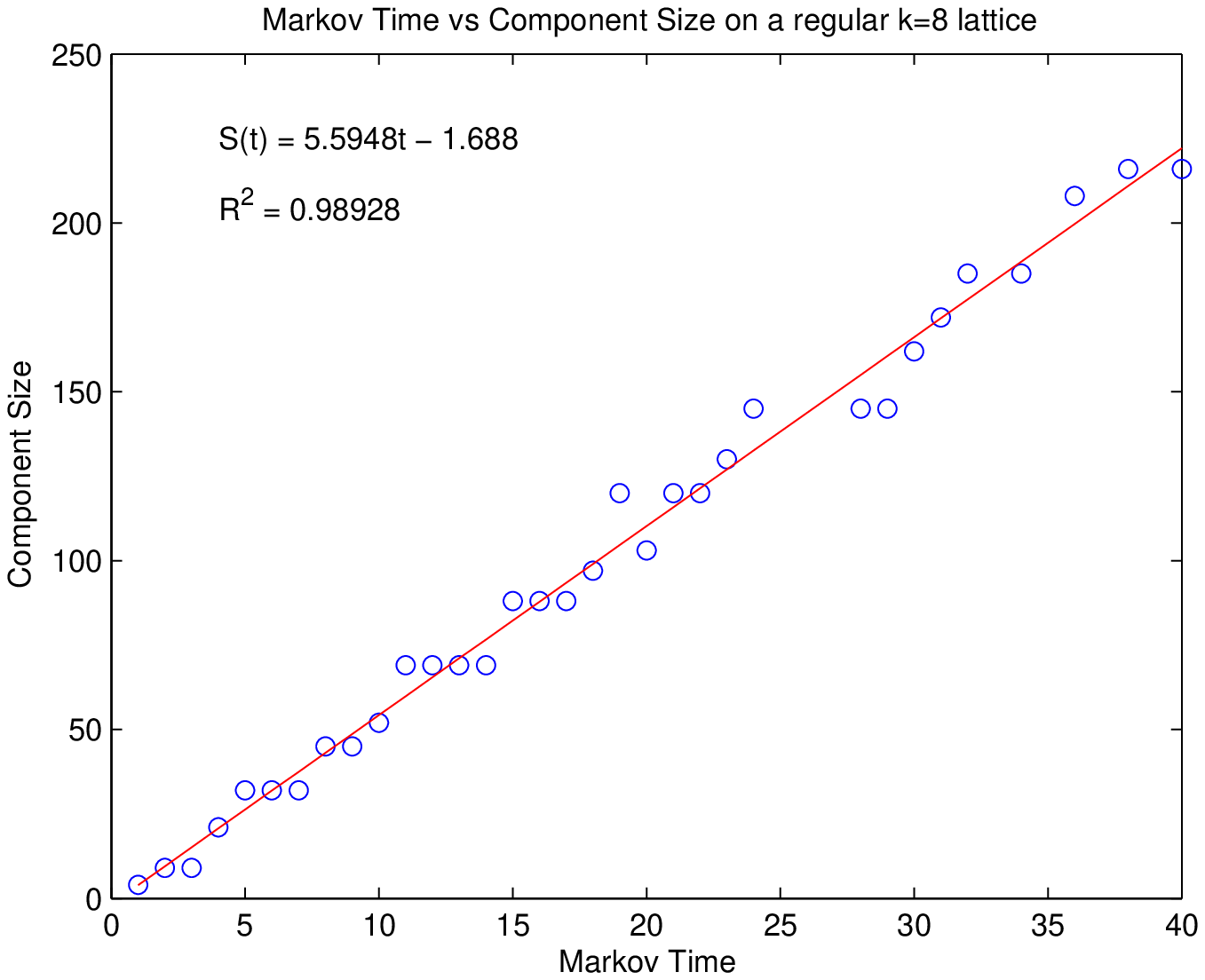}  
  \includegraphics[width=1\columnwidth]{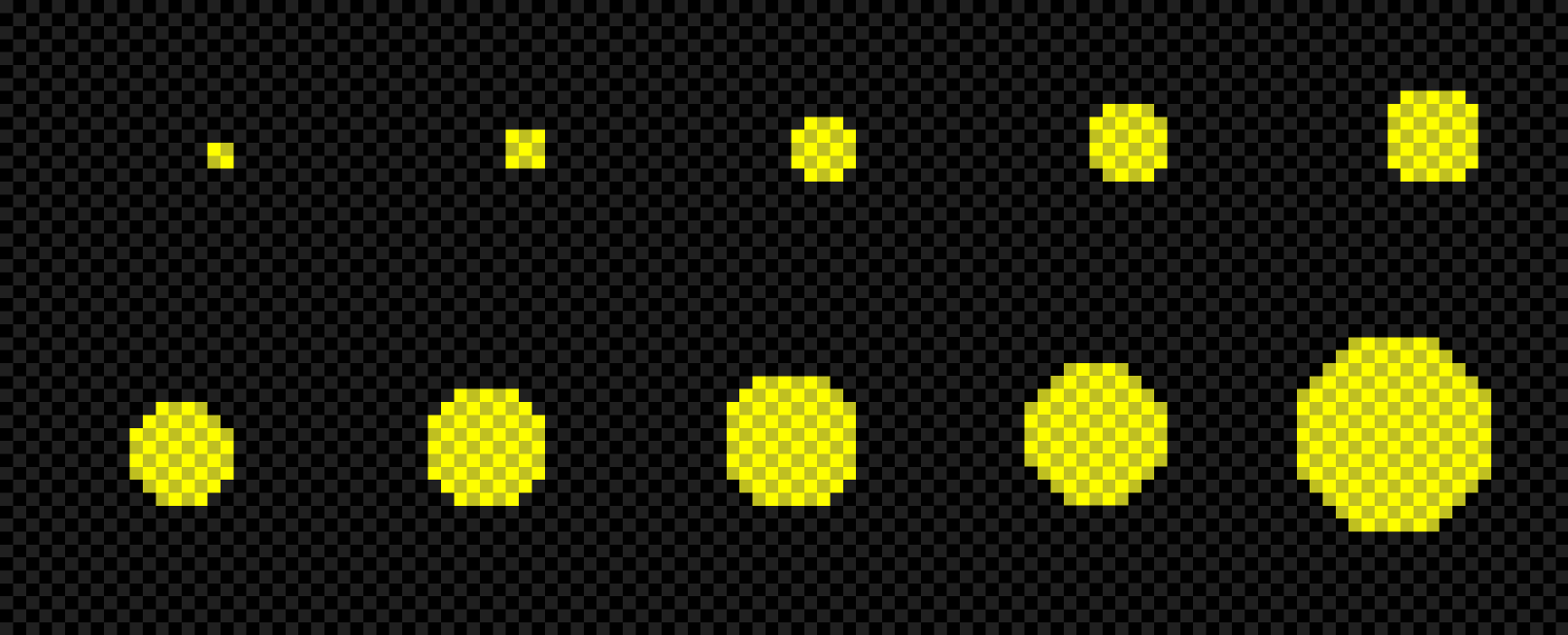}
\caption[Square lattice diffusion]{(top) Correlation of Markov time with size on an square lattice. (bottom) The transient components found by severability on a regular lattice at Markov times $t = \{1,2, 3, 5, 8, 10,11,15,18,32  \}$. Each block is connected to all eight of its neighbouring blocks by a single undirected edge of weight $1$.}
\label{fig:square}
\end{center}
\end{figure}

\clearpage
\onecolumngrid

\section{Ring of small-worlds: commutativity \& locality}
We further explore commutativity as in Figure \ref{fig:images}, by looking at a ring of small-world networks and comparing against OSLOM.
We first generate small-world networks using the Watts-Strogatz model.
Each node is first connected to its 2 neighbors on both sides.
Then every edge is rewired with independent probability $0.1$, but such that multi-edges cannot exist, so a small world with a total of 5 nodes will not be rewired from a 5-clique.

Note that whereas severability gives the same results when looking at a single small-world network compared to a ring of four of them, OSLOM does not.
Some of this is equivalent behaviour, as OSLOM chooses to not consider the entire network as a valid community.
For the small-worlds of size 5, 10, and 20, OSLOM returns all individual nodes, which is as valid of an answer as the entire network.
However, for the largest of the small-world networks of size 40, OSLOM chooses to split it up into 3 pieces, which is not what it chose in the ring of 4 small-worlds.
Severability always gives the small-world at the appropriate times, as it is truly local.

Additionally, OSLOM demonstrates trouble when the scales of the networks are very different.
It is unable to recover the 5-clique of the smallest small-world, despite the 5-clique being recoverable when the other communities are of the same size.
This comes from the imposition of the same resolution on all communities implicit in OSLOM.
\label{sec:smallworlds}
\begin{figure}[tbp]
\begin{center}
    \includegraphics[width=0.8\columnwidth]{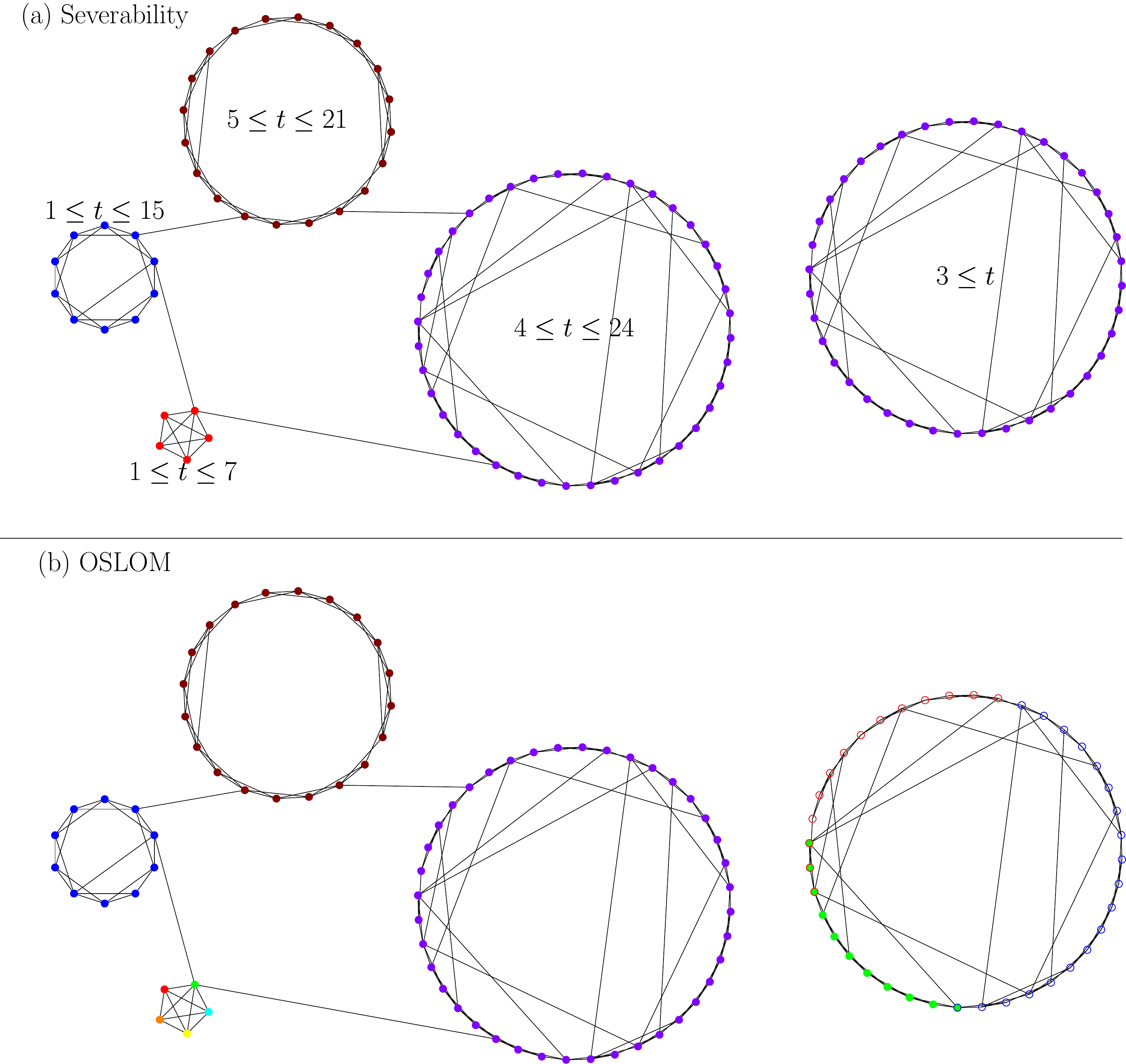}
    \caption[Ring of smallworlds]{Ring of small world networks of sizes 5, 10, 20, and 40 generated using the Watts-Strogatz model. (a) At different times, severability recovers each small-world network, as expected. Additionally, when the largest small-world is in isolation, it still correctly recovers it a component. (b) OSLOM recovers the three larger small-worlds, but splits up the smallest one, even though in other experiments it can recover 5-cliques. Additionally, when the largest small-world is given in isolation, OSLOM breaks it up, giving three overlapping communities instead. }
\label{fig:smallworlds}
\end{center}
\end{figure}

\clearpage
\section{Co-existence of different timescales}

\label{sec:coexistence}
\begin{figure}[tbp]
\begin{center}
    \includegraphics[width=0.8\columnwidth]{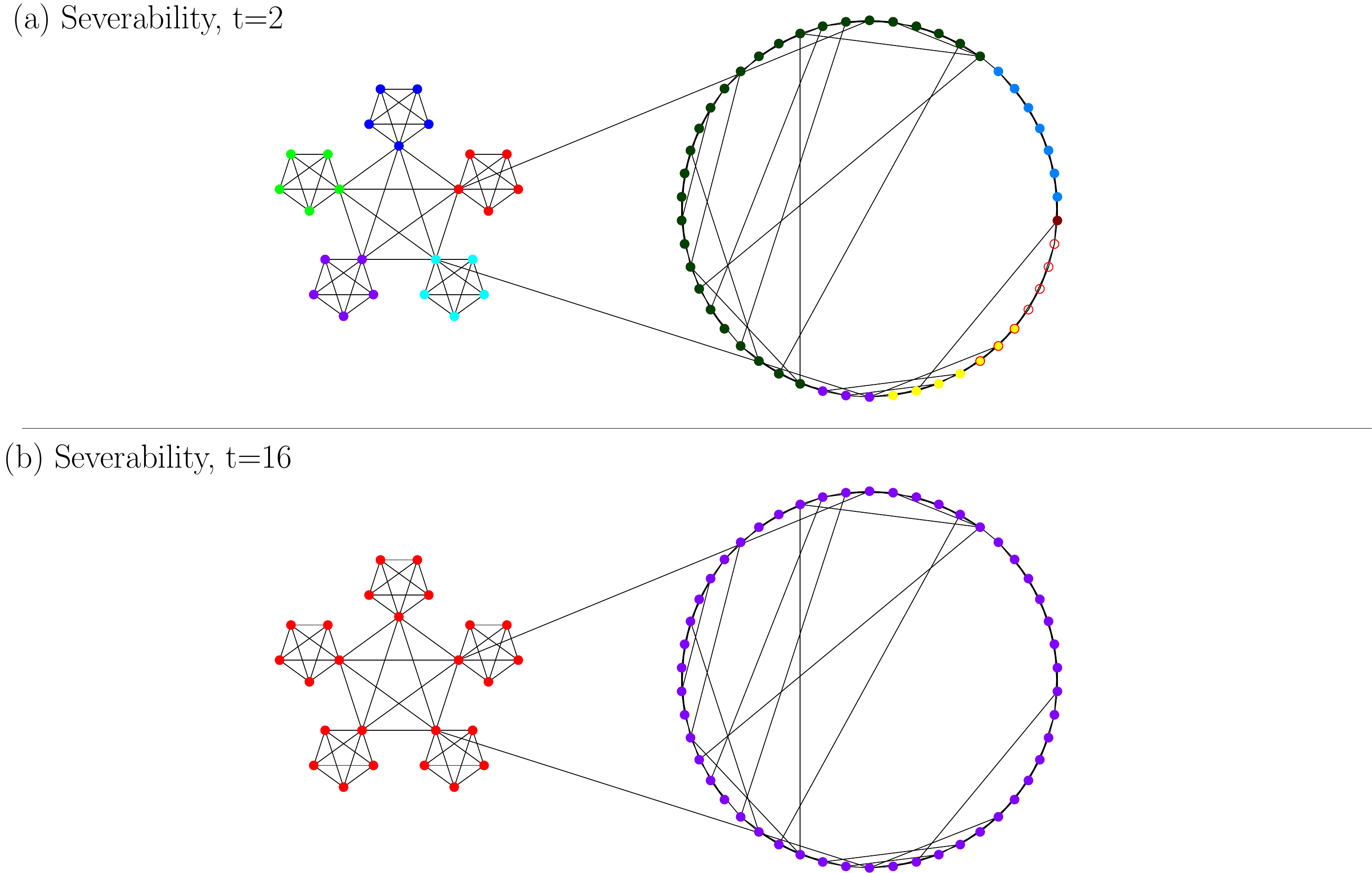}
    \caption[Two timescales]{5-clique of 5-cliques attached to a small world network of size 50 generated using the Watts-Strogatz model. (above) At Markov time 2, the 5-cliques are recovered, but not the 50-small world. At Markov time 16, the size 50 small world is recovered, but the 5-cliques aggregate into a 5-clique of 5-cliques. At no one time are both the 5-cliques and the 50 small world simultaneously recovered, because they exist on different time scales.}
\label{fig:5star}
\end{center}
\end{figure}

\clearpage

\section{Word Association Extended}
\label{sec:word_extended}
Figure~\ref{fig:word_associate} only depicted the components including ``nature'' and the orphans directly connected to that word. However, this is only a small snippet of the entire network. Here, we have displayed all the other components that have at least one link to ``nature'', but do not include the word itself. As with Figure~\ref{fig:word_associate}, the maximum search size $S=50$ and the Markov time $t=2$.

  \begin{center}
\includegraphics[width=0.9\textwidth]{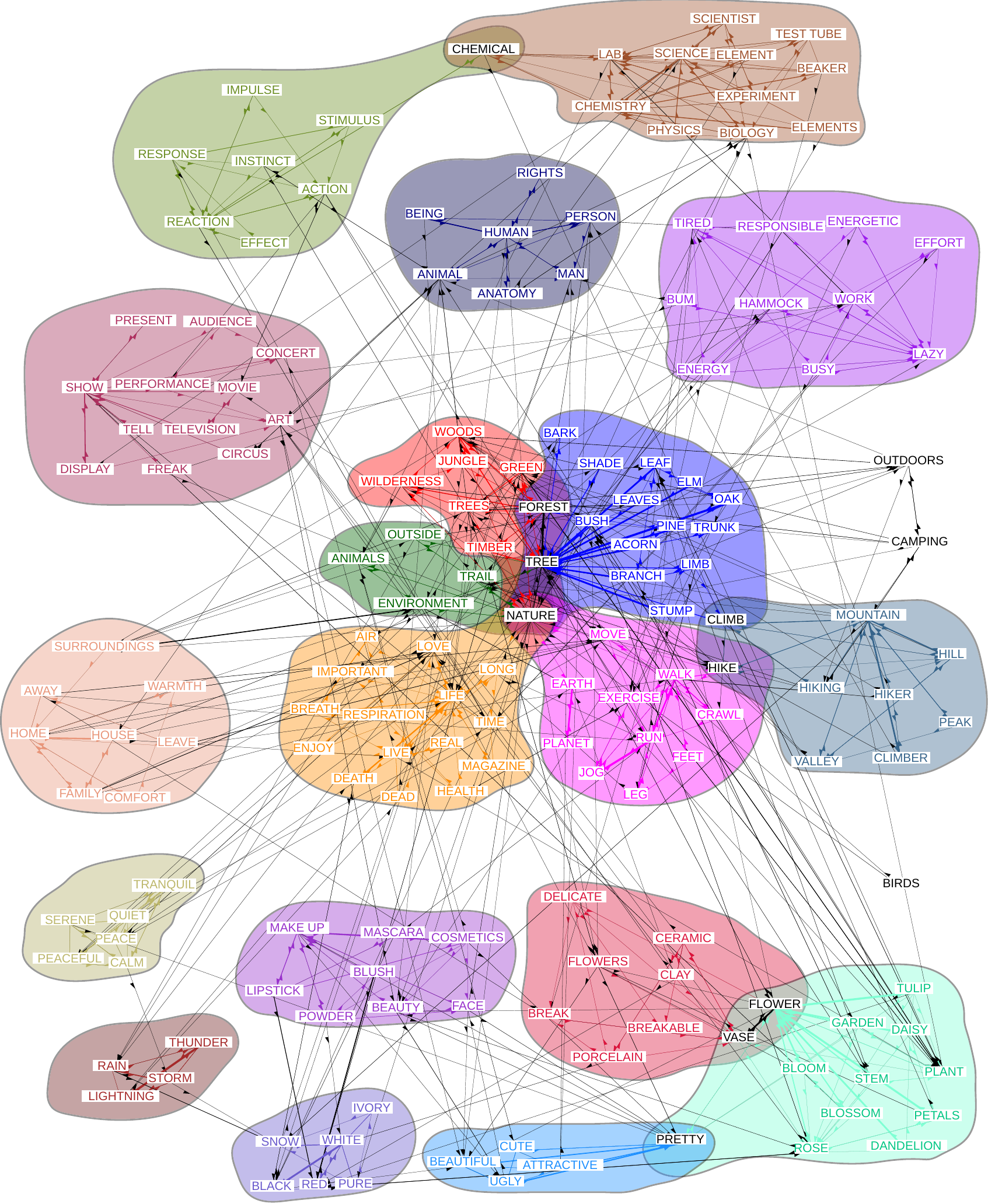}
  \end{center}

\clearpage
\twocolumngrid

\end{document}